\documentclass[conference]{IEEEtran}

\usepackage{amssymb, amsmath}
\usepackage[shortcuts]{extdash}
\usepackage{listings}
\usepackage{array}
\usepackage{relsize}
\usepackage{calculator}
\usepackage{pgfplots}
\usepackage{tikz}
\usetikzlibrary{positioning, shapes, decorations.markings, arrows, calc, patterns}
\usepackage{color, soul}
\usepackage{marginnote}
\usepackage{etoolbox}
\usepackage{multirow}
\usepackage{algorithm}
\usepackage{algpseudocode}
\usepackage{wasysym}
\usepackage{ifthen}
\usepackage{xstring}
\usepackage{array,booktabs,calc}
\usepackage{ragged2e}
\usepackage{todonotes}
\usepackage{graphicx}
\usepackage[caption=false]{subfig}
\usepackage{wrapfig}
\usepackage[inline]{enumitem}
\usepackage{url}
\usepackage{macros}
\iftrue % color
	\graphicspath{{img-color/}}
\else % grayscale
	\selectcolormodel{gray}
	\graphicspath{{img-gray/}}
\fi

\iftrue % comments
	\newcommand{\comment}[2]{\todo[backgroundcolor=#1!25, linecolor=#1!50!black, bordercolor=#1!50!black]{#2}}
	\newcommand{\slcomment}[1]{\comment{green}{#1  SL}}
	\newcommand{\pgcomment}[1]{\comment{red}{#1  PG}}
	\newcommand{\uhcomment}[1]{\comment{yellow}{#1  UH}}
	\newcommand{\cdcomment}[1]{\comment{blue}{#1  CD}}
\else % no comments
	\newcommand{\comment}[2]{}
	\newcommand{\slcomment}[1]{}
	\newcommand{\pgcomment}[1]{}
	\newcommand{\uhcomment}[1]{}
	\newcommand{\cdcomment}[1]{}
\fi

\newcommand{\node}{\Lambda}

\newcommand{\pert}{\Delta}%\mathrm{pert}}
\newcommand{\refr}{\mathrm{ref}}
\newcommand{\synch}{\mathit{synch}}

\newcommand{\states}{\Gamma}
\newcommand{\fstates}{\states^{\mathsf{F}}}
\newcommand{\nfstates}{\states^{\mathsf{NF}}}
\newcommand{\system}{\mcs}

\newcommand{\psystem}{\system}
\newcommand{\psystemfull}{\psystem = (\pert, N ,T, R, \epsilon, \mu)}
\newcommand{\nextstates}{\mathit{next}}
\newcommand{\usynch}{U}
\newcommand{\ustates}{\states_{\usynch}}

\renewcommand{\th}[1]{$#1^{\mathrm{th}}$}

\newcommand{\init}{\state_{0}}

\newcommand{\failvecs}{\mcf}
\newcommand{\fire}{\mathit{fire}}
\newcommand{\update}{\mathit{update}}

\newcommand{\Phase}{\Phi}
\newcommand{\PhaseTwo}{\Psi}
\newcommand{\state}{\sigma}
\newcommand{\statefull}{\state = \langle k_1, \ldots, k_T \rangle}
\newcommand{\fvec}{F}
\newcommand{\trans}{\rightarrow}
\newcommand{\transskip}{\twoheadrightarrow}
\newcommand{\sucbase}{\mathrm{succ}}
\newcommand{\suc}{\overset{\text{\scriptsize$\trans$}}{\sucbase}}
\newcommand{\sucskip}{\overset{\text{\scriptsize$\transskip$}}{\sucbase}}
\newcommand{\ptf}{\tau}
\newcommand{\phiupdate}{\mcu}
\newcommand{\pfail}{\mathsf{PMF}}
\newcommand{\pfailvec}{\mathsf{PFV}}

\newcommand{\orderp}{\mathsf{PCF}}%{\eta}
\newcommand{\tocomplex}{p^\mathbb{C}}

\newcommand{\meanphasepos}{\overline{\Phase}}

\newcommand{\phasecon}{\theta}
\newcommand{\Phasecon}{\Theta}
\newcommand{\conffull}{\langle \phasecon_1, \ldots, \phasecon_N \rangle}

\newcommand{\ampdraw}{I}
\newcommand{\ampdrawi}{\ampdraw_{I}}
\newcommand{\ampdrawt}{\ampdraw_{T}}
\newcommand{\ampdrawr}{\ampdraw_{R}}
\newcommand{\watthours}{W}
\newcommand{\watthoursi}{\watthours_{I}}
\newcommand{\watthourst}{\watthours_{T}}
\newcommand{\watthoursr}{\watthours_{R}}

\newcommand{\power}{\mathrm{pow}}
\newcommand{\powerskip}{\overset{\text{\scriptsize$\transskip$}}{\power}}

\newcounter{cnt}

\definecolor{darkgrey}{rgb}{0.3,0.3,0.3}
\definecolor{greylight}{rgb}{0.7,0.7,0.7}
\definecolor{greyblue}{rgb}{0.0,0.0,0.5}
\definecolor{greyred}{rgb}{0.5,0.0,0.0}
\definecolor{lightgreyblue}{rgb}{0.5,0.5,0.9}
\definecolor{lightgreyred}{rgb}{0.6,0.2,0.2}
\definecolor{lightgreyorange}{rgb}{0.7,0.5,0.2}

\newcommand{\runningexampleparams}
{
	\def \n {10}
	\def \radius {1.27cm}
	\def \rotation {0}
	\def \margin {10}
	\def \nodeminsize {0.45cm}
	\def \ldistance {0.6cm}
    \def \tmargin {0}
    \def \refracbound {3}
	\def \nodeedgeradius {{\radius - \nodeminsize / 2}}
}

\newcommand{\drawopacity}{1}
\newcommand{\fireandrefracopacity}{0.35}
\newcommand{\drawlabelcolour}{black}
\newcommand{\drawstatelabels}{true}
\newcommand{\nodecolour}{darkgrey}
\newcommand{\lcolour}{black}
\newcommand{\firenodecolour}{lightgreyred}
\newcommand{\refracnodecolour}{lightgreyblue}

\newcommand{\statelabeltextsize}{\small}

\newcommand{\drawnodeandarc}[9]
{
    \node (statelabel) [] {$\statel$};   	
    
    \ifnum #8 = 0
        \ifnum #7 = -1
            \node (#7) [opacity = \drawopacity, draw, circle, color = \nodecolour, double, double distance = 2pt,
                minimum size = #5] at ({360/#4 * (#7 - 1) + #2}:#1) {};
        \else
            \ifnum #7 = \n
            	\node [rotate = 0, opacity = \fireandrefracopacity, scale = 0.85, fill = \firenodecolour, star, star points=5, star point ratio = 1.6] at ({360/#4 * (#7 - 1) + #2}:#1) {};
                \node (#7) [opacity = \drawopacity, draw, color = \firenodecolour, circle, minimum size = #5]
                    at ({360/#4 * (#7 - 1) + #2}:#1) {};
            \else
				\ifnum #7 < #9
					\draw [fill = \refracnodecolour, opacity = \fireandrefracopacity] ({360/#4 * (#7 - 1) + #2}:#1) circle [radius={#5/2}];
					\node (#7) [opacity = \drawopacity, draw, color = \refracnodecolour, circle, minimum size = #5] at ({360/#4 * (#7 - 1) + #2}:#1) {};
				\else
					\node (#7) [opacity = \drawopacity, draw, color = \nodecolour, circle, minimum size = #5] at ({360/#4 * (#7 - 1) + #2}:#1) {};
				\fi
            \fi
        \fi
        \def \nodetext{}
    \else
        \ifnum #7 = -1
            \node (#7) [opacity = \drawopacity, draw, color = \nodecolour, inner sep = 0pt, circle, double, fill,
                double distance = 2pt, minimum size = #5]
                at ({360/#4 * (#7 - 1) + #2}:#1) {\color{white}{\statelabeltextsize{#8}}};
        \else
            \ifnum #7 = \n
            	\node [rotate = 0, opacity = 1.0, scale = 0.85, fill = \firenodecolour, star, star points=5, star point ratio = 1.6] at ({360/#4 * (#7 - 1) + #2}:#1) {};            
                \node (#7) [opacity = \drawopacity, draw, inner sep = 0pt, color = \firenodecolour, circle, 
                    minimum size = #5]
                    at ({360/#4 * (#7 - 1) + #2}:#1) {\color{white}{\statelabeltextsize{#8}}};
            \else
            	\ifnum #7 < #9
	                \node (#7) [opacity = \drawopacity, draw, inner sep = 0pt, color = \refracnodecolour, circle, fill, minimum size = #5] at ({360/#4 * (#7 - 1) + #2}:#1) {\color{white}{\statelabeltextsize{#8}}};
	           	\else
	                \node (#7) [opacity = \drawopacity, draw, inner sep = 0pt, color = \nodecolour, circle,  minimum size = #5] at ({360/#4 * (#7 - 1) + #2}:#1) {{\statelabeltextsize{#8}}};            	
            	\fi
            \fi
        \fi
        \def \nodetext{#8}
    \fi

    \draw[opacity = \drawopacity, ->, >=stealth, line width=0.65pt] ({360/#4 * (#7 - 1) + #2 + #3}:#1)
        arc ({360/#4 * (#7 - 1) + #2 + #3}:{360/#4 * (#7) + #2 - #3}:#1);

    \IfStrEq{#7}{2}
    {
    \draw[opacity = \drawopacity, -, dotted, line width=0.65pt] ({360/#4 * (#7 - 1) + #2 + #3}:{#1 + 0.35cm})
        arc ({360/#4 * (#7 - 1) + #2 + #3}:{360/#4 * (#7) + #2 - #3}:#1);    	
    }
        
    \IfStrEq{\drawstatelabels}{true}
    {
    	\IfStrEq{#7}{1}
    	{
	    	\node (#7{label}) [color = \lcolour] at ({360/#4 * (#7 - 1) + #2}:{#1 + #6}) {\statelabeltextsize{\color{\drawlabelcolour}{$k_{#7}$}}};    	
    	}
    	\IfStrEq{#7}{2}
    	{
	    	\node (#7{label}) [color = \lcolour] at ({360/#4 * (#7 - 1) + #2}:{#1 + #6}) {\statelabeltextsize{\color{\drawlabelcolour}{$k_{#7}$}}};    	
    	}
    }
}

\newcommand{\drawnodeandarch}[9]
{
    \ifnum #8 = 0
        \ifnum #7 = -1
            \node (#7) [opacity = \drawopacity, draw, circle, color = \nodecolour, double, double distance = 2pt,
                minimum size = #5] at ({360/#4 * (#7 - 1) + #2}:#1) {};
        \else
			\ifnum #7 < #9
				\draw [fill = \refracnodecolour, opacity = \fireandrefracopacity] ({360/#4 * (#7 - 1) + #2}:#1) circle [radius=0.3cm];
				\node (#7) [opacity = \drawopacity, draw, color = \refracnodecolour, circle, minimum size = #5] at ({360/#4 * (#7 - 1) + #2}:#1) {};
			\else
				\node (#7) [opacity = \drawopacity, draw, color = lightgray, circle, minimum size = #5] at ({360/#4 * (#7 - 1) + #2}:#1) {};
			\fi
        \fi
        \def \nodetext{}
    \else
        \ifnum #7 = -1
            \node (#7) [opacity = \drawopacity, draw, color = \nodecolour, inner sep = 0pt, circle, double, fill,
                double distance = 2pt, minimum size = #5]
                at ({360/#4 * (#7 - 1) + #2}:#1) {\color{white}{\statelabeltextsize{#8}}};
        \else
           	\ifnum #7 < #9
				\node (#7) [opacity = \drawopacity, draw, inner sep = 0pt, color = \refracnodecolour, circle, fill, minimum size = #5] at ({360/#4 * (#7 - 1) + #2}:#1) {\color{white}{\statelabeltextsize{#8}}};
			\else
				\node (#7) [thick, opacity = \drawopacity, draw, inner sep = 0pt, color = \nodecolour, circle, double, minimum size = #5] at ({360/#4 * (#7 - 1) + #2}:#1) {};%\color{gray}{\statelabeltextsize{#8}}};
			\fi
        \fi
        \def \nodetext{#8}
    \fi

    \draw[opacity = \drawopacity, -triangle 45, color = lightgray] ({360/#4 * (#7 - 1) + #2 + #3}:#1)
        arc ({360/#4 * (#7 - 1) + #2 + #3}:{360/#4 * (#7) + #2 - #3}:#1);

    \IfStrEq{\drawstatelabels}{true}
    {
    \node [opacity = \drawopacity, color = \lcolour,
        label={[label distance=(#1 + #6)]{360/#4 * (#7 - 1) + #2}:\scriptsize{\color{\drawlabelcolour}{\ifnum #8 > 0 $\tocomplex(#7)$ \fi}}}] {}; %k_#7{=}#8$
    }
}

\tikzset
{
    ->-/.style =
    {
        decoration =
        {
            markings,
            mark = at position 0.5 with {\arrow[scale = 1.5]{>}},
        },
        postaction={decorate}
    }
}
        
\newcommand{\drawtransitionarc}[8]
{
    \def \arccolour {orange}
    
    \draw[opacity = #8, densely dotted, color = \arccolour] ({360 / #4 * (#5 - 1) + #2}:#7)
        -- ({360 / #4 * (#5 - 1) + #2}:#1);

    \draw[opacity = #8, densely dotted, color = \arccolour] ({360 / #4 * (#5 - 1 + #6) + #2}:#7)
        -- ({360 / #4 * (#5 - 1 + #6) + #2}:#1);
        
    \draw[opacity = #8, ->-, color = \arccolour] ({360/#4 * (#5 - 1) + #2}:#7)
        arc ({360 / #4 * (#5 - 1) + #2}:{360 / #4 * (#5 - 1 + #6) + #2}:#7);
}

\newcommand{\dnach}[2]
{
    \drawnodeandarch{\radius}{\rotation}{\margin}{\n}{\nodeminsize}{\ldistance}{#1}{#2}{\refracbound};
}

\newcommand{\dnac}[2]
{
    \drawnodeandarc{\radius}{\rotation}{\margin}{\n}{\nodeminsize}{\ldistance}{#1}{#2}{\refracbound};
}

\newcommand{\dta}[4]
{
    \drawtransitionarc{\nodeedgeradius}{\rotation}{\margin}{\n}{#1}{#2}{#3}{#4};
}

\begin{document}

%\title{The Power of Flashing Fireflies: \\Formal Analysis of Power Consumption in Networks of Pulse-Coupled Oscillators}
\title{The Power of Synchronisation: \\Formal Analysis of Power Consumption in Networks of Pulse-Coupled Oscillators}

\author{
	\IEEEauthorblockN{Paul Gainer, Sven Linker, Clare Dixon, Ullrich Hustadt, and Michael Fisher}
	\IEEEauthorblockA{Department of Computer Science, University of Liverpool \\
		Liverpool, L69 3BX, United Kingdom
}}

\maketitle

\begin{abstract}
% We introduce a methodology to analyse the power consumption of
% biologically-inspired synchronisation protocols based on
% formal methods. To that end, we present a general
% model for the well-known formalism of pulse-coupled oscillators, 
% which can be instantiated with different parameters, e.g., the
% length of the refractory period and the coupling strength. We use 
% the model-checker \prism{} to analyse different instantiations
% of our model and discuss trade-offs with respect to 
% power consumption and time to synchronise. \slcomment{prelim. abstract.}{}

%We assess the power consumption of network synchronisation
%   protocols, particularly the energy required to synchronise all
%   nodes across a network. We use the widely adopted approach of
%   bio-inspired, pulse-coupled oscillators to achieve network-wide
%   synchronisation and provide an extended formal model of just such a
%   protocol, enhanced with structures for recording energy
%   usage. Exhaustive analysis is then carried out through formal
%   verification, utilising the PRISM model-checker to calculate the
%   resources consumed on each possible system execution. This allows
%   us to assess a range of parameter instantiations and to explore
%   trade-offs between power consumption and time to synchronise. This
%   provides a principled basis for the formal analysis of a much
%   broader range of large-scale network protocols.

Nature-inspired synchronisation protocols have been widely adopted
to achieve consensus within wireless sensor networks. We assess
the power consumption of such
protocols, particularly the energy required to synchronise all
nodes across a network. We use the widely adopted model of
bio-inspired, pulse-coupled oscillators to achieve network-wide
synchronisation and provide an extended formal model of just such a
protocol, enhanced with structures for recording energy
usage. Exhaustive analysis is then carried out through formal
verification, utilising the PRISM model-checker to calculate the
resources consumed on each possible system execution. This allows
us to assess a range of parameter instantiations and to explore
trade-offs between power consumption and time to synchronise. This
provides a principled basis for the formal analysis of a much
broader range of large-scale network protocols.
\end{abstract}

\IEEEpeerreviewmaketitle

\section{Introduction}
\label{sec:introduction}

%\paragraph{INTRODUCTORY TEXT - SYNCHRONISATION, SYNCHRONISATION OF WSNS, POWER CONSUMPTION - WHY DO WE CARE?}

%\paragraph{POWER CONSUMPTION IN WSNS}

Minimising power consumption is a critical design consideration for wireless sensor
networks (WSNs)~\cite{rhee2004techniques, albers2010energy}.
Once deployed a WSN is generally expected to function
independently for long periods of time. In particular,
regular battery replacement can be costly and impractical for remote sensing applications.
Hence, it is of the utmost importance to reduce the power consumption of the 
individual nodes by choosing low-power hardware and/or energy efficient
protocols. However, to make informed choices,
it is also necessary to have
good estimations of the power consumption
for individual nodes. While the general power consumption of the hardware can be extracted from
data sheets, estimating the overall power consumption of different protocols is more demanding.

Surveys conducted by Irani and Pruhs~\cite{irani2005algorithmic}
and Albers~\cite{albers2010energy} investigated algorithmic problems
in power management, in particular power-down mechanisms at the system and
device level.
Soua and
Minet provided a general taxonomy for the analysis of wireless network protocols with respect to
energy efficiency~ \cite{soua2011survey} by identifying the 
contributing factors of energy wastage, for instance packet collisions and unnecessary idling.  
%More specifically,
%Oller et al. analysed whether \emph{wake-up radio} based medium-access control protocols are
%more energy efficient than \emph{duty-cycle} based protocols~\cite{oller2016has}.
These detrimental effects can be overcome by allocating time slots for
communication between nodes. That is, nodes within a network
synchronise their clock values and use different time slots for communication
to avoid packet collisions~\cite{yick2008survey, rhee2009clock}.

A number of biologically inspired protocols for synchronisation have been
proposed~\cite{taniguchi2005distributed,tyrrell2006fireflies,bojic2014scalability,lipa2015firefly}
and have been shown to be robust with
respect to the topology of the network~\cite{werner2005firefly}. 
They are well-suited for WSNs since centralised control is not required
to achieve synchrony.
The protocols build on the underlying mathematical
model of pulse-coupled oscillators (PCOs); integrate-and-fire
oscillators with pulsatile coupling, such that when an oscillator fires it
induces some phase-shift response %(usually positive)
determined by a \emph{phase response function}. Over time the
mutual interactions can lead to all oscillators firing synchronously.
The PCO synchronisation model we employ was first proposed by
Peskin~\cite{peskin1975mathematical}
%to encapsulate the behaviour of the cardiac pacemaker. This model was
and later extended by Mirollo and Strogatz~\cite{mirollo1990synchronization}
who proved that %more than
several oscillators with the same frequency would always
synchronise under the assumption of a fully coupled network.
Later work by Lucarelli and Wang showed that
this assumption could be relaxed, by proving that oscillators would always achieve
synchrony if the coupling graph of the network was connected~\cite{lucarelli2004decentralized}.

Simulating such a system %is time-efficient, and 
provides good estimates
for its typical behaviour, but may exclude corner cases %extreme cases 
where some unexpected behaviour is exhibited.
%and occasional erratic behavior.
To mitigate against this, 
we analyse the energy-consumption for the synchronisation of a network of
PCOs using formal methods. Instead of using simulations, we use
\emph{probabilistic model checking}~\cite{kwiatkowska2007stochastic}
to exhaustively examine all possible runs of the system. 
Probabilistic model checking can be used to formally specify performance
measures and to analyse trade-offs in Markovian models~\cite{baier2010performability,baier2014trade}.
Using this technique we can calculate expected mean and worst-case energy costs for a
network. 

\begin{wrapfigure}{l}{.45\linewidth}
	\includegraphics[width=\linewidth]{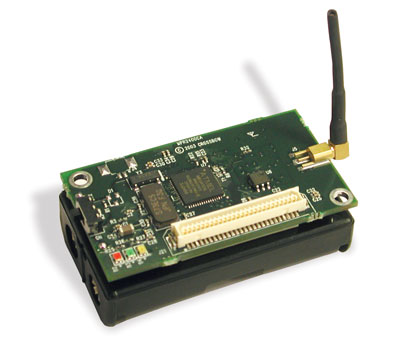}
	\caption{The MICAz wireless measurement system.}
        \label{fig:micaz}
\end{wrapfigure}
In this work we abstract away from the modelling of individual oscillators
and use a
\emph{population model}~\cite{emerson1999asymmetry,donaldson2006symmetry,gainer2016probabilistic,gainer2017investigating}
to encode information about groups of oscillators sharing the same configuration.
Furthermore, we introduce \emph{broadcast failures} where %, i.e.,
an oscillator may fail to broadcast its message. Since WSNs operate in stochastic
environments under uncertainty we encode these failures within a
\emph{probabilistic} model.
Our model also encapsulates means to associate different current draws with its states, thus enabling
us to measure the energy consumption of the overall network. We employ the
probabilistic model checker \prism{} \cite{kwiatkowska2011prism} to analyse the 
average and worst-case energy consumption for both the synchronisation of arbitrarily configured
networks, and restabilisation of a network, where a subset of oscillators desynchronised.
To that end, we instantiate the model to analyse the power consumption of the MICAz wireless measurement system (see Fig.~\ref{fig:micaz}).

%\cite{bojic2014scalability}
%\cite{lipa2015firefly}
%\cite{taniguchi2005distributed}
%\cite{tyrrell2006fireflies}
% PCO with ref. periods.
%\cite{degesys2008synchronization}

%\paragraph{FORMAL METHODS APPLIED TO WSN PROTOCOLS}
%Formal methods have been used ...
%\cite{kamali2015formal}
%%\cite{fehnker2006formal}
%\cite{gallina2013process} (FORMAL!!!)
%\cite{heidarian2012analysis}

%\paragraph{FORMAL (PROBABILISTIC) METHODS APPLIED TO ANALYSE WSN PROTOCOLS}
%P.MODEL CHECKING: \cite{kwiatkowska2007stochastic}
%%PCTL: \cite{hansson1994logic}
%TOOLS: \cite{katoen2011ins}, , \cite{hahn2014iscasmc}
%POPULATION MODELS: \cite{gainer2016probabilistic}, \cite{gainer2017investigating}, NEED MORE??

%\paragraph{FORMAL METHODS APPLIED TO ANALYSE SYNCHRONISATION PROTOCOLS }
%\cite{bartocci2010detecting}
%\cite{pfeifer1999formal}
%\cite{gainer2017investigating}
%\cite{heidarian2012analysis}

%\paragraph{MICAz MOTE}

Exact time synchronisation, where all clocks always agree on
their value, is never achieved for real-world deployments of synchronising devices~\cite{bojic2014scalability}.
Hardware imperfections result in different clock frequencies, environmental
factors influence radio transmission, and network congestion leads to package collisions
and loss~\cite{hull2004mitigating}. Consequently, the precision of
synchronisation is not required to be exact, and it is sufficient for all
oscillators to fire within some defined time window~\cite{bojic2014scalability}.
The size of this window depends on the application. Some applications may
require a very small window, for instance distributed sensing of mobile objects,
while others may prefer energy efficiency at the cost of synchronisation precision~\cite{rhee2009clock}.
To this end we extend the binary notion of synchronisation discussed in \cite{gainer2017investigating} by defining a metric derived from the complex order parameter of Kuramoto \cite{kuramoto2012chemical,kuramoto1975self}
that captures the degree of synchrony of a fully connected network of
oscillators as a real value in the interval $[0, 1]$.

The structure of the paper is as follows.
In Sect.~\ref{sec:relatedwork} we discuss related work, and in
Sect.~\ref{sec:oscillatormodel} we introduce the general PCO model, from which 
we derive population models in Sect.~\ref{sec:populationmodel}. Section~\ref{sec:sync_metric}
introduces the derived synchronisation metric. % in the context of population models.
The construction of the formal model used for the analysis is presented
in Sect.~\ref{sec:model}. Subsequently, in Sect.~\ref{sec:eval}  we evaluate
the results for certain parameter instantiations and discuss their trade-offs 
with respect to power consumption and time to synchronise. Section~\ref{sec:conc} concludes
the paper.

\section{Related Work}
\label{sec:relatedwork}

While formal methods, in particular model checking, have been
successfully used to model and analyse protocols
for wireless sensor systems,
the number of possible configurations that needs to be considered for larger
WSNs impacts their feasibility. Chen et al. reviewed how different
formal methods may be used to investigate ad-hoc routing protocols \cite{chen2013review},
suggesting that model checking is suitable for small networks, while analytical methods
%must be employed 
are necessary for larger networks.
Yue and Katoen~\cite{yue2010leader} used probabilistic model checking to
optimise the energy consumption of a leader election protocol in networks
of up to nine nodes.
Probabilistic model checking was also used by 
Fehnker and Gao \cite{fehnker2006formal} to analyse flooding and gossiping protocols
in networks of up to eight nodes, 
%Model checking was used to analyse networks of up to eight nodes,
%While they were able to use model checking to analyse networks
%with at most eight nodes,
however it was necessary to use Monte Carlo simulations for the analysis
of larger networks. H{\"o}fner and Kamali~\cite{hofner2013quantitative}
used statistical model checking, an
approach combining model checking, Monte Carlo sampling and hypothesis testing,
to analyse a routing protocol for a network of sixteen nodes.
Heidarian et al. used model checking to analyse 
clock synchronisation for
medium access protocols \cite{heidarian2012analysis}. They considered both fully-connected networks and 
line topologies with up to four nodes. 
Model checking of biologically inspired coupled oscillators has
also been investigated by Bartocci et al. \cite{bartocci2010detecting}. They present
a subclass of timed automata \cite{alur1994theory} %whose mutual interactions were
suitable to model biological oscillators,
and an algorithm to detect synchronisation properties. However, their analysis
was restricted to a network of three oscillators.  
%whose mutual interactionsthat interact, which gradually influence each other

% THIS SOUNDS LIKE THE ABSTRACT?
In \cite{gainer2017investigating} we introduced a %preliminary
formal population model for a network of PCOs, 
%and analysed the emergence of synchrony in the network for different sets of parameters.
and investigated the expected time to achieve synchronisation and the probability 
%with which we would expect 
for an arbitrarily configured population of oscillators to synchronise. 
In our model the oscillators synchronise over a finite set of discrete clock values, and
the oscillation cycle includes a \emph{refractory period}  %The refractory period
 at the start of the oscillation cycle where an oscillator cannot be
perturbed by other firing oscillators. 
%In addition to helping to achieve synchrony the
%refractory period also 
This corresponds to a period of time where a WSN node enters a
low-power idling mode.
%, we informally discussed the power consumption of the network.
In this work we extend this approach by introducing a metric for global power consumption and discuss refinements of the model that allows us to formally reason about much larger populations of oscillators. 

Wang et al. proposed an energy-efficient strategy for
the synchronisation of PCOs \cite{wang2012energy}. 
In contrast to our work, they consider real-valued clocks
and \emph{delay-advance phase response}
functions, where both positive and negative phase shifts can occur.
%That is, if an oscillator is perturbed within the first half of
%its oscillation cycle the perturbation is negative, while in the second
%half of the cycle the perturbation is positive.
A result of their choice of phase response function is that synchronisation time is independent of
the length of the refractory period,
in contrast to our model.
Furthermore, they assume that the initial phase difference between
oscillators has an upper bound. They achieve
synchrony for refractory periods larger than half the cycle,
while our models do not always synchronise in these cases, as
%for such refractory periods.
we do not impose a bound on the phase difference of the oscillators. We 
consider all possible differences in phase since we examine the energy consumption for
the resynchronisation of a subset of oscillators.

Konishi and Kokame conducted an analysis of PCOs where a perceived
pulse immediately resets the oscillators to the start of their cycle \cite{konishi2008synchronization}.
Their goal was to maximise refractory period length,
while still achieving synchronisation within some number of clock cycles. Similarly
to our work, they restricted their analysis to a fully coupled network. %to be within an all-to-all coupling.
They assumed that the synchronisation protocol was implemented as part of the physical
layer of the network stack by using capacitors to generate pulses, therefore 
their clocks were continuous and had different frequencies. 
We assume that the synchronisation protocol resides on a higher layer,
where the clock values are discretised and oscillate with the same frequency.

The energy consumption of the MICAz mote varies with the mode of its
RF transceiver. The node has a receive mode, three transmission modes, and two low power idling
modes. While Kramer and Geraldy~\cite{kramer2006energy} conducted an empirical investigation into
the energy consumption of the \mbox{MICAz}, with respect to the different modes,
Webster et al.~\cite{webster2017performance} used probabilistic model checking to formally analyse very small numbers (not populations)
of MICAz nodes, particularly the effect of clock drift on synchronisation.

%presented a formal analysis using probabilistic model-checking of very small numbers (not populations)
%of MICAz nodes, particularly the effect of clock drift on synchronisation.

%%% Local Variables: 
%%% mode: latex
%%% TeX-master: "main"
%%% End: 
\section{Discrete Oscillator Dynamics}
\label{sec:oscillatormodel}

% which we denote by blah blah
We consider a fully-coupled network of PCOs %pulse-coupled oscillators
with identical dynamics over discrete time.
%We denote the set of these oscillators by $\network = \{1, 2, \ldots\}$,
%where each $i \in \network$ corresponds to a single pulse-coupled
%oscillator.
The \emph{phase} of an oscillator $i$ 
%$\network$
at time $t$ is denoted by $\phi_i(t)$.
The phase of an oscillator progresses through a sequence of discrete
integer values bounded by some $T \ge 1$.
The phase progression over time of a single uncoupled oscillator
is determined by the successor function, where the phase
increases over time until it equals $T$, at which point the
oscillator will fire in the next moment in time and the phase
will reset to one. The phase progression of an uncoupled
oscillator is therefore cyclic with period $T$, and we refer to
one cycle as an \emph{oscillation cycle}.

When an oscillator fires, its firing may not be perceived
by any of the other oscillators coupled to it. We call this a
\emph{broadcast failure} and denote its probability by $\mu\in [0,1]$.
Note that $\mu$ is a global parameter, hence the chance of broadcast
failure is identical for all oscillators.
When an oscillator fires, and a broadcast failure does not occur,
it perturbs the phase of all oscillators to which it is coupled;
we use $\alpha_i(t)$ to denote the number of all other
oscillators %in $\network$
that are coupled to $i$ and will fire
at time $t$.
%\begin{definition}
The \emph{phase response function} is a positive increasing function
$\pert : \{1, \ldots, T\} \times \mathbb{N} \times \mathbb{R}^+
\to \mathbb{N}$
%with $\pert(\Phi, \alpha, \epsilon) \ge 0$ for all
%$\Phi, \alpha, \epsilon$,
%$\langle \Phi, \alpha, \epsilon \rangle \in \{1, \ldots, T\} \times \mathbb{N} \times \mathbb{R}^+$,
that maps the phase of an oscillator $i$,
the number of other oscillators perceived to be firing by $i$,
and a real value defining the strength of the coupling between oscillators, to an integer value corresponding to the perturbation to phase
induced by the firing of oscillators where broadcast failures did
not occur.
%\label{def:pert}
%\end{definition}

We can introduce a refractory period into the oscillation cycle
of each oscillator. 
A refractory period is an interval of discrete values
$[1, R] \subseteq [1, T]$ where $1 \le R \le T$ is the size of the
refractory period, such that if $\phi_i(t)$ is inside the interval, for some
oscillator $i$ at time $t$, then $i$ cannot be perturbed by other
oscillators to which it is coupled.
If $R = 0$ then we set $[1, R] = \emptyset$, and there is no refractory period at all.
%\begin{definition}
The \emph{refractory function}
$\refr : \{1, \ldots, T\} \times \mathbb{N} \to \mathbb{N}$
is defined as $\refr(\Phi, \delta) = 0$ if $\Phi \in [0, R]$,
or $\refr(\Phi,\delta) = \delta$ otherwise, and
takes as parameters $\delta$, the degree of perturbance to the phase of
an oscillator, and $\phi$, the phase,
and returns zero if $\phi$ is in the refractory period defined by $R$,
or $\delta$ otherwise.
%\label{def:ref}
%\end{definition}

We now introduce the \emph{update function}
and \emph{firing predicate}, which respectively denote the updated
phase of an oscillator $i$ at time $t$ in the
next moment in time, and the firing of oscillator $i$ at time $t$,
\begin{IEEEeqnarray}{rl}
	\update_i(t) & = 1 + \refr(\phi_i(t), \pert(\phi_i(t), \alpha_i(t), \epsilon)) \\
	\fire_i(t) & = \update_i(t) > T.	
\end{IEEEeqnarray}
The %precise 
phase evolution %of the phase of %an 
of an oscillator $i$ 
over time is
%then 
given by
\begin{equation}
\phi_i(t+1) =
	\begin{cases}
		1 & \text{if } \fire_i(t) \\
		\update_i(t) & \text{otherwise}.
	\end{cases}
\end{equation}

\section{Population Model}
\label{sec:populationmodel}

%\pgcomment{The evol function has been removed, we simply use the successor function
%for standalone phase evolution. The pert function is renamed to 
%\emph{phase response function} (now $\pert$) to more closely match the
%literature (see Phase Response Curve).}{}
Let $\pert$ be a phase response function for a network
of $N$ identical oscillators, where each oscillator is coupled to all other
oscillators, and where the coupling strength is given by the constant
$\epsilon$. Each oscillator has a phase in $1, \ldots, T$, and a
refractory period defined by $R$. The probability of broadcast failure
in the network is $\mu \in [0, 1]$. We define a \emph{population model}
of the network as $\psystemfull$. Oscillators in our model have
identical dynamics, and two oscillators are indistinguishable if they
share the same phase. We therefore encode the global state of the model
as a tuple $\langle k_1, \ldots, k_T \rangle$
where each $k_\Phase$ is the number of oscillators with
phase $\Phase$.
%\begin{definition}

\begin{figure}[b]
\begin{center}
\begin{tabular}{cc}
%$\state_0$ & $\state_1$ & $\state_2$ \\
\begin{tikzpicture}
	\runningexampleparams
	\def \statel {\state_0}
	\dnac{1}{2};\dnac{2}{1};\dnac{3}{0};\dnac{4}{0};\dnac{5}{5};
	\dnac{6}{0};\dnac{7}{0};\dnac{8}{0};\dnac{9}{0};\dnac{10}{0};
\end{tikzpicture} &
\begin{tikzpicture}
	\runningexampleparams
	\def \statel {\state_1}	
	\dnac{1}{0};\dnac{2}{0};\dnac{3}{0};\dnac{4}{0};\dnac{5}{0};
	\dnac{6}{2};\dnac{7}{1};\dnac{8}{0};\dnac{9}{0};\dnac{10}{5};
	\dta{1}{5}{0.8cm}{1};	
	\dta{2}{5}{0.6cm}{0.8};	
	\dta{5}{5}{0.4cm}{0.6};	
\end{tikzpicture} \\
\begin{tikzpicture}
	\runningexampleparams
	\def \statel {\state_2}	
	\dnac{1}{6};\dnac{2}{0};\dnac{3}{0};\dnac{4}{0};\dnac{5}{0};
	\dnac{6}{0};\dnac{7}{0};\dnac{8}{0};\dnac{9}{0};\dnac{10}{2};
	\dta{6}{4}{0.8cm}{1};
	\dta{7}{4}{0.6cm}{0.8};
	\dta{10}{1}{0.4cm}{0.6};
\end{tikzpicture} &
\begin{tikzpicture}
	\runningexampleparams
	\def \statel {\state_3}	
	\dnac{1}{2};\dnac{2}{6};\dnac{3}{0};\dnac{4}{0};\dnac{5}{0};
	\dnac{6}{0};\dnac{7}{0};\dnac{8}{0};\dnac{9}{0};\dnac{10}{0};
	\dta{10}{1}{0.8cm}{0.8};
	\dta{1}{1}{0.6cm}{0.6};
\end{tikzpicture} \\
\end{tabular}
\caption{Evolution of the global state over four discrete time steps.}
\label{fig:example1}
\end{center}
\end{figure}
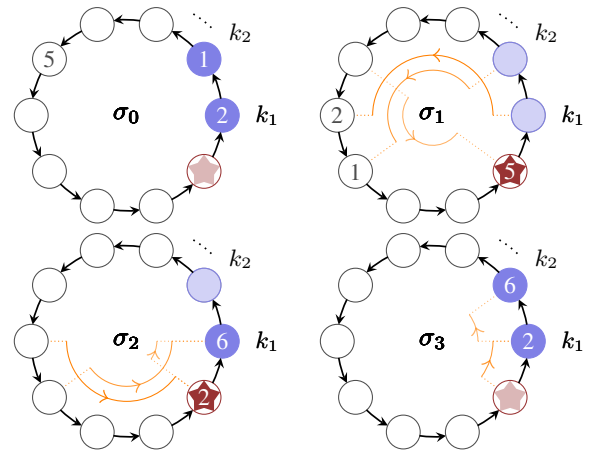

A \emph{global state} of $\psystem$ %of a population model $\psystemfull$
is a $T$-tuple
$\state \in \{0, \ldots, N\}^T$, where
$\state = \langle k_1, \ldots, k_T\rangle$
and $\sum_{\Phase = 1}^T k_\Phase = N$.
%We denote 
We denote by $\states(\psystem)$ the set of all global states of $\psystem$,
and will simply use $\states$ when $\psystem$ is clear from the context.
Fig.~\ref{fig:example1} shows four global states of a population
model of $N = 8$ oscillators with $T = 10$ discrete values for their phase
and a refractory period of length $R = 2$.
For example \mbox{$\state_0 = \langle 2, 1, 0, 0, 5, 0, 0, 0, 0, 0 \rangle$}
is the global state where two oscillators have a phase of one,
one oscillator has a phase of two, and five oscillators have a phase of five.
The starred node indicates the number of oscillators with phase 
ten that will fire in the next moment in time, while the shaded nodes
indicate oscillators with phases that lie within the refractory period
(one and two).
If no oscillators have some phase $\Phase$ then we omit the $0$ in the
corresponding node.

We distinguish between states where one or more oscillators are about
to fire, and states where no oscillators will fire at all. We refer
to these states as \emph{firing states} and \emph{non-firing} states
respectively.
%\begin{definition}
Given a population model $\psystem$, a global state
$\langle k_1, \ldots, k_T\rangle \in \states$
is a \emph{firing state} if, and only if, $k_T > 0$.
We denote by $\fstates(\psystem)$ the set of all firing states of $\psystem$,
and denote by $\nfstates(\psystem)$ the set of all non-firing states of $\psystem$.
Again we will simply use $\fstates$ or $\nfstates$ when $\psystem$ is clear from the context.
%\label{def:firingstates}
%\end{definition}

\subsection{Successor States}
We now define how the global state of a population model evolves over
time. Since our population model encodes uncertainty in the form
of broadcast failures, firing states may have more than one possible
successor state.
We denote the transition from a firing state
$\state$ to a possible successor state $\state^\prime$ by
$\state \trans \state^\prime$.
With every firing state $\state \in \fstates$ we associate a non-empty set
of \emph{failure vectors}, where each failure vector is a tuple
of broadcast failures that could occur in $\state$.
A \emph{failure vector} is a $T$-tuple
where the \th{\Phase} element denotes the number of broadcast failures that occur for all
oscillators with phase $\Phase$.
If the \th{\Phase} element is $\star$ then
no oscillators with a phase of $\Phase$ fired. %, for all $1 \le \Phase < T$.
We denote the set of all possible failure vectors by $\failvecs$.
Oscillators with phase less than $T$ may fire due to
being perturbed by the firing of oscillators with a phase of $T$. This
is discussed in detail later in this section.\footnote{Due to space limitations we refer the reader to~\cite{gainer2017investigating}
for a detailed description of how the set of all possible failure vectors for a
firing state can be constructed.}

\paragraph{Non-Firing States}
A non-firing state will always have exactly one successor state, as
there is no oscillator
that is about to fire. Therefore, the dynamics of all oscillators in that
state are determined solely by the successor function. That is, the phase of
every oscillator is simply updated by one in the next time step.
This continues until one or more oscillators fire
and perturb the phase of other oscillators. 
Given a sequence of global states
$\state_0, \state_1, \ldots, \state_{n-1}, \state_n$
where $\state_0, \ldots \state_{n-1} \in \nfstates$
and $\state_n \in \fstates$, we omit transitions between
$\state_i$ and $\state_{i + 1}$ for $0 \le i < n$, and instead introduce a direct
transition $\state_0 \transskip \state_n$ from the first non-firing state to the next
firing state in the sequence.
This is a refinement of the model presented
in~\cite{gainer2017investigating}, as
omitting these intermediate transitions results in smaller models. While
the state space remains the same the number of transitions in the model
is substantially decreased.
Hence the time and resources required to check desirable properties are
reduced.
%Hence a reduction in the 
We denote the transition from a non-firing
state $\state$ to its single successor state $\state^\prime$ by
$\state \transskip \state^\prime$.
%\slcomment{Mention that $\delta$ is always less than $T$}{}
%For a non-firing state $\state = \langle k_1, \ldots, k_T \rangle \in \nfstates$
%let $\delta$ be the highest phase of any oscillator in $\state$,
%i.e.,
%$\delta = \max \{ \Phase \mid \Phase \in \{1, \ldots  T\} \land k_\Phase > 0\}$.
%The \emph{successor function} $\sucskip : \nfstates \to \fstates$
%mapping a non-firing state to its successor firing state is then given by
%$\sucskip(\langle k_1, \ldots, k_T\rangle) = {\{0\}^{T-\delta}}^\frown \langle k_1, \ldots, k_{\delta} \rangle $, %where ${}^\frown$ indicates vector concatenation.
For example, in Fig.~\ref{fig:example1} state $\state_0$ is a non-firing state,
%where $\delta = 5$
and its successor $\sucskip(\state_0) = \state_1$
is a firing state where all oscillator phases have been \mbox{increased by $5$}.

\paragraph{Encoding Chain Reactions}
For real deployments of protocols for synchronisation the effect of one
or more oscillators firing may cause other oscillators to which they are
coupled to fire in turn. This may then cause further
oscillators to fire, and so forth, and we refer to this event as a
\emph{chain reaction}. When a chain reaction occurs it can lead to
multiple groups of oscillators being triggered to fire and being
\emph{absorbed} by the initial group of firing oscillators. 

These chain reactions are usually near-instantaneous events.
Since we model the oscillation cycle as a progression through a number
of discrete states, we choose to encode chain reactions by updating the
phases of all perturbed oscillators in a single time step. 
Since we only consider fully-connected topologies, any oscillators
sharing the same phase will always perceive the same number of other
oscillators firing.

For the global state $\state_1$ of Fig.~\ref{fig:example1} we can
see that five oscillators will fire in the next moment in time.
In the successive state $\state_2$, the single oscillator with a phase
of seven in $\state_1$ perceives the firing of the five oscillators.
The induced perturbation causes the single oscillator to also fire and
therefore be absorbed by the group of five. The remaining two
oscillators with a phase of six in $\state_1$ perceive six oscillators
to be firing, but the induced perturbation is insufficient to
cause them to also fire, and they instead update their phases to
ten.

%For each phase $\Phase \in \{1, \ldots, T\}$
%we define the function $\alpha^\Phase\colon \states \times \failvecs\to \{1,\dots,N\}$,
%where $\alpha^\Phase(\state, \fvec)$ is the number of 
%oscillators with a phase greater than $\Phase$ perceived to be firing by
%oscillators with phase $\Phase$ in a firing state, %incorporating
%given that the broadcast failures defined in the failure vector $\fvec$ occur.
%Given a firing state $\state = \langle k_1, \ldots, k_T \rangle \in \fstates$
%and a failure vector $\fvec = \langle f_1, \ldots, f_T \rangle$,
%we can calculate $\alpha^1(\state, \fvec),\ldots,\alpha^T(\state, \fvec)$
%using the functions defined iteratively as follows:\slcomment{write this in English?}
%\begin{IEEEeqnarray}{rll}
%	\alpha^\Phase(\state, \fvec){} & = &
%	\begin{cases}
%		0 & \text{if } \Phase{=}T\\
%		\renewcommand{\arraystretch}{1}
%		\begin{array}[t]{@{}l@{}}
%		\alpha^{\Phase{+}1}(\state{,}\fvec){+} \\
%		k_{\Phase{+}1}{-}f{\vphantom{f^{\Phi}}}_{\Phase{+}1}		
%		\end{array}		
%		& \text{if} \ 		\renewcommand{\arraystretch}{1}
%		\begin{array}[t]{@{}l@{}}
%		\Phase{<}T , \thinspace f_{\Phase{+}1}{\ne}\star, \\
%		\text{and} \thinspace \fire^{\Phase{+}1}(\state{,}\fvec) \\		
%		\end{array} \\		
%		\alpha^{\Phase{+}1}(\state{,}\fvec) & \text{otherwise}\\
%	\end{cases} \\
%	\fire^\Phase(\state, \fvec) & = & \update^\Phase(\state, \fvec) > T \\
%	\update^\Phase(\state, \fvec) & = & 1 + \refr(\Phase, \pert(\Phase, \alpha^\Phase(\state, \fvec), \epsilon)).
%\end{IEEEeqnarray}

\paragraph{Firing States}
With every firing state we have by definition that at least one oscillator
is about to fire in the next time step. Since the firing of this oscillator
may, or may not, result in a broadcast failure we can see that
at least two failure vectors will be associated with any firing state,
and that additional failure vectors will be associated with firing states
where more than one oscillator is about to fire.
Given a firing state $\state$ and a failure vector $\fvec$ associated with
that state, we can compute the successor of $\state$.
For each phase $\Phase \in \{1{,}\ldots{,}T\}$ we calculate the
number of oscillators with a phase greater than $\Phase$  perceived
to be firing by oscillators with phase $\Phase$. We simultaneously
calculate $\update^\Phase(\state, \fvec)$, the updated phase of
oscillators with phase $\Phase$, and $\fire^\Phase(\state, \fvec)$, the
predicate indicating whether or not oscillators with phase $\Phase$ fired.
Details of these constructions are given in~\cite{gainer2017investigating}.

%We first define the following functions:
%$\alpha^\Phase$ calculates the number of oscillators with a phase
%greater than $\Phase$ perceived to be firing by oscillators with
%phase $\Phase$, $\update^\Phase$ calculates the updated phases
%of oscillators with phase $\Phase$, and $\fire^\Phase$ 
%determines if oscillators with phase $\Phase$ have fired. 

We can then define the function that maps phase values to their updated
values in the next moment in time. Since we do not distinguish
between oscillators with the same phase we only calculate a single
updated value for their phase.
%\begin{definition}
The \emph{phase transition function}
$\ptf : \fstates \times \{1{,} \ldots{,} T\} \times \failvecs \to \mathbb{N}$
maps a firing state $\state$, a phase $\Phase$, and a failure
vector $\fvec$ for $\state$, to the updated phase in the next moment in time,
with respect to the broadcast failures defined in $\fvec$, and is defined as %\label{def:globalstates}
%\end{definition}

%\begin{figure}[b]
%\begin{tabular}{c@{}c@{}c}
%$\state_0$ & $\state_1$ & $\state_2$ \\
%\begin{tikzpicture}
%	\runningexampleparams
%	\dnac{1}{0};\dnac{2}{1};\dnac{3}{0};
%	\dnac{4}{2};\dnac{5}{0};\dnac{6}{0};
%\end{tikzpicture} &
%\begin{tikzpicture}
%	\runningexampleparams
%	\dnac{1}{0};\dnac{2}{0};\dnac{3}{0};
%	\dnac{4}{1};\dnac{5}{0};\dnac{6}{2};
%\end{tikzpicture} &
%\begin{tikzpicture}
%	\runningexampleparams
%	\dnac{1}{3};\dnac{2}{0};\dnac{3}{0};
%	\dnac{4}{0};\dnac{5}{0};\dnac{6}{0};
%\end{tikzpicture} \\
%\end{tabular}
%\caption{Evolution of the global state over three discrete time steps.}
%\label{fig:example1}We would have to run
%\end{figure}ed as
\begin{equation}
	\ptf(\state, \Phase, \fvec) =
	\begin{cases}
		1 & \text{if } \fire^\Phase(\state, \fvec) \\
		\update^\Phase(\state, \fvec) & \text{otherwise}. \\
	\end{cases}
\end{equation}
%\label{def:ptf}
%\end{definition}
%\begin{lemma}
%The range of the function $\tau$ is bound by $T$. That is, for any
%$\pi$, for any possible failure vector $B$ for $\pi$,  and for all
%$\Phi \in \{1, \ldots, T\}$, we have that $1 \le \tau(\pi, \Phi, B) \le T$.
%\label{lem:taubound}
%\end{lemma}
%\begin{proof}
%See Appendix~\ref{app:proof_taubound}.
%\end{proof}

Let $\phiupdate_\Phase(\state, \fvec)$ be the set of phase values $\PhaseTwo$
where all oscillators with phase $\PhaseTwo$ in $\state$ will have the updated phase
$\Phase$ in the next time step, with respect to the broadcast failures defined in $\fvec$.
Formally,
$\phiupdate_\Phase(\state, \fvec) = \{\PhaseTwo \mid \Phase \in \{1,\ldots,T\} \land
		\ptf(\state, \PhaseTwo, \fvec) = \Phase\}$.
We can now calculate the successor state of a firing state $\state$ and define how the
model evolves over time.
Observe that the population model does not encode oscillators leaving or joining the
network, therefore the population $N$ remains constant. 
%\begin{definition}
The \emph{firing successor function}
$\suc : \fstates \times \failvecs \to \states$ maps
a firing state $\state$ and a failure vector $\fvec$ to a global state $\state^\prime$,
and is defined as
$\suc(\langle k_1, \ldots, k_T \rangle, \fvec) = \langle k_1^\prime, \ldots, k_T^\prime \rangle$, 
where $k_\Phase^\prime{=}\sum_{\PhaseTwo \in \phiupdate_\Phase(\state, \fvec)} k_\Phase$ for $1 \le \Phase \le T$.
%\label{def:fsucc}
%\end{definition}

%\begin{lemma}
%The number of oscillators is invariant during transitions, i.e., the successor
%function only creates tuples that are states of the given model. Formally,
%let \(\pi= \langle n_1, \ldots, n_T \rangle\) and \(\pi^\prime = 
%\langle n_1^\prime, \ldots, n_T^\prime \rangle\) be two states
%of a model \(\psystem\) such that \(\pi^\prime = \suc(\pi, B)\), where $B$
%is some possible failure vector for \(\pi\). Then
%$\sum_{\Phi = 1}^T n_\Phi = \sum_{\Phi = 1}^T n_\Phi^\prime = N.$
%\label{lem:nochange}
%\end{lemma}
%\begin{proof}
%See Appendix~\ref{app:proof_nochange}.
%\end{proof}

\subsection{Transition Probabilities}
We now define the probabilities that will label the transitions in our model.
Given a global state $\state \in \states$,
if $\state$ is a non-firing state then it has exactly one successor state.
If $\state$ is a firing state then to construct the set of possible
successor states we must first construct  $\failvecs_{\state}$, the set of all
possible failure vectors for $\state$. 
%\begin{definition}
Given a global state $\state \in \states$ we define
$\nextstates(\state)$,
the set of all successor states of $\state$, as
\begin{equation}
\nextstates(\state) =
\begin{cases}
\{\suc(\state, \fvec) \mid \fvec \in \failvecs_{\state}\} & \text{if } \state \in \fstates \\
\{\sucskip(\state)\} & \text{if } \state \in \nfstates.
\end{cases}
\end{equation}
%\pgcomment{Maybe just get rid of this paragraph?}{}
%Note that for a firing state $\state \in \fstates$ it may be the case that
%for some $\fvec_1, \fvec_2 \in \failvecs_\state$ with $\fvec_1{\ne}\fvec_2$,
%we have that $\suc(\state, \fvec_1){=}\suc(\state, \fvec_2)$, and hence
%that $|\nextstates(\state)|{<}|\failvecs_{\state}|$.
%\label{def:next}
%\end{definition}

For every non-firing state $\state \in \nfstates$ we have 
\mbox{$|\nextstates(\state)| = 1$}, since there is always exactly one
successor state $\sucskip(\state)$, and we label the transition
$\state \transskip \sucskip(\state)$ with probability one.
We now consider each firing state
\mbox{$\state = \langle k_1, \ldots, k_n \rangle \in \fstates$},
and for every successor $\suc(\state, \fvec) \in \nextstates(\state)$,
we calculate the probability that will label \mbox{$\state \trans \suc(\state, \fvec)$}.
Recalling that $\mu$ is the probability of a broadcast failure
occurring, let \mbox{$\pfail : \{1{,}\ldots{,}N\}^2 \rightarrow [0, 1]$}
be a probability mass function where
\mbox{$\pfail(k, f) =  \mu^{f} (1{-}\mu)^{k - f} \binom{k}{f}$}
is the probability that $f$ broadcast failures occur given that
$k$ oscillators fire. Then let
$\pfailvec : \fstates \times \failvecs \to [0, 1]$ be the function
mapping a firing state $\state = \langle k_1, \ldots, k_T \rangle$
and a failure vector
$\fvec = \langle f_1, \ldots, f_T \rangle\in \failvecs$
to the probability of the failures in $\fvec$ occurring in $\state$, given by 
\begin{equation}
\textstyle\pfailvec(\state, \fvec) =
		\prod_{\Phase = 1}^T
		\begin{cases}		
			\pfail(k_\Phase , f_\Phase) & \text{if} \ f_\Phi \ne \star \\
			1  & \text{otherwise}.
		\end{cases}
\end{equation}

We can now describe the evolution of the global state over time. 
%\uhcomment{This is the same definition of QEST2017 paper. A problem
 %with it is that it does not mention probabilities.}{}
A \emph{run} of a population model $\psystem$ is an infinite
sequence $\state_0, \state_1, \state_2, \cdots$,
where $\state_0$ is called the \emph{initial state}, and
$\state_{i + 1}  \in \nextstates(\state_i)$ for all $i \ge 0$.
%We denote the set of all possible runs of $\psystem$ by
%$\run(\psystem)$. 

\section{Synchronisation and Metrics}
\label{sec:sync_metric}
Given a population model $\psystemfull$, and a global state
$\state \in \states$, we say that $\state$ is
\emph{synchronised} %, denoted by $\synch(\state)$,
if all oscillators in $\state$ share the same phase.
%Formally, a global state $\state = \langle k_1, \ldots, k_T \rangle$
%is \emph{synchronised} if, and only if, there is some
%$\Phase \in \{1, \ldots, T\}$ such that $k_\Phase = N$, and
%for all $\PhaseTwo \ne \Phase$ we have that $k_\PhaseTwo = 0$.
We say that a run of the model 
$\state_0, \state_1, \state_2, \cdots$ \emph{synchronises} if there
exists an $i > 0$ such that $\state_i$ is synchronised.
%exists 
%For any run $\run = \state_0, \state_1, \state_2, \ldots \in \run(\psystem)$,
%if there exists an $i > 0$ such that $\synch(\state_i)$ then
%we say that $\run$ \emph{synchronises}.
Note that if a state $\state_i$ is synchronised
then any successor state $\state_{i + 1}$ of $\state_i$ will also be synchronised.
The population model does not encode oscillators leaving or joining the
network, therefore the population $N$ remains constant. 
That is, the global state remains synchronised forever.

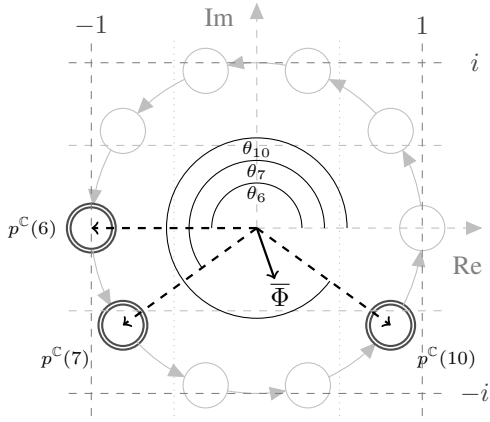
\begin{figure}[bt]
\begin{center}
\begin{tikzpicture}
	\def \n {10}
	\def \radius {2.2cm}
	\def \nodeedgeradius {1.55cm}
	\def \lradius {1.4cm}
	\def \rotation {0}
	\def \margin {7.7}
	\def \nodeminsize {0.6cm}
	\def \ldistance {0.12cm}
    \def \tmargin {0}
    \def \refracbound {0}
	\dnach{1}{0};\dnach{2}{0};\dnach{3}{0};\dnach{4}{0};\dnach{5}{0};
	\dnach{6}{2};\dnach{7}{1};\dnach{8}{0};\dnach{9}{0};\dnach{10}{4};
	
	\draw[dashed, color = gray] (-2.5, \radius)--(2.5, \radius);
	\draw[dashed, color = gray] (-2.5, -\radius)--(2.5, -\radius);
	\draw[dashed, color = lightgray] (-2.5, 1.1)--(2.5, 1.1);
	\draw[dashed, color = lightgray] (-2.5, -1.1)--(2.5, -1.1);
	\draw[dashed, color = gray] (\radius, -2.5)--(\radius, 2.5);
	\draw[dashed, color = gray] (-\radius, -2.5)--(-\radius, 2.5);
	\draw[dotted, color = lightgray] (1.1, -2.5)--(1.1, 2.5);
	\draw[dotted, color = lightgray] (-1.1, -2.5)--(-1.1, 2.5);
	
	\tikzset{->-/.style = {decoration = {markings,
		mark = at position #1 with {\arrow{>}}}, postaction = {decorate}}}
	
	\coordinate (origin) at (0, 0);
	\coordinate[] (iaxisend) at (0, 3.0);
	\node[color = gray] (imlabel) at (-0.5, 2.8) {Im}; 
	\coordinate[] (raxisend) at (3.0, 0);
	\node[color = gray] (relabel) at (2.8, -0.5) {Re};
	\draw[dashed, -triangle 45, color = lightgray] (origin) to (iaxisend);
	\draw[dashed, -triangle 45, color = lightgray] (origin) to (raxisend);

%	\filldraw[pattern = horizontal lines, pattern color = lightgray] (origin)++(1.2, 0) arc (0:270:1.2) -- (origin) -- (0.6, 0.0);
%	\filldraw[fill = white] (origin)++(0.9, 0) arc (0:180:0.9) -- (origin) -- (0.6, 0.0);
%	\filldraw[pattern = crosshatch dots, pattern color = lightgray] (origin)++(0.9, 0) arc (0:180:0.9) -- (origin) -- (0.6, 0.0);
%	\filldraw[fill = white] (origin)++(0.6, 0) arc (0:45:0.6) -- (origin) -- (0.6, 0.0);
%	\filldraw[pattern = north west lines, pattern color = lightgray] (origin)++(0.6, 0) arc (0:45:0.6) -- (origin) -- (0.6, 0.0);

	\draw[color = black] (origin)++(0.6, 0) arc (0:180:0.6);
	\draw[color = black] (origin)++(0.9, 0) arc (0:216:0.9);
	\draw[color = black] (origin)++(1.2, 0) arc (0:324:1.2);
%	\draw[color = black, -open triangle 45] (origin)++(0.6, 0) arc (0:180:0.6);
%	\draw[color = black, -open triangle 45] (origin)++(0.9, 0) arc (0:216:0.9);
%	\draw[color = black, -open triangle 45] (origin)++(1.2, 0) arc (0:324:1.2);
	
%	\draw let \p1 = (2) in (\x1, 0)-- (2);
%	\draw let \p1 = (2) in (origin)--(\x1, 0);
	\draw[dashed, ->, line width = 0.3mm] let \p1 = (6) in  (origin) to (\x1, \y1);
	\draw[dashed, ->, line width = 0.3mm] let \p1 = (7) in  (origin) to (\x1, \y1);
	\draw[dashed, ->, line width = 0.3mm] let \p1 = (10) in  (origin) to (\x1, \y1);
%	\node[color = black] (theta0) at (0, 0.25) {$\scriptstyle{\theta_{6}}$};
%	\node[color = black] (theta1) at (-0.6, -0.2) {$\scriptstyle{\theta_{7}}$};
%	\node[color = black] (theta2) at (-0.5, -0.8) {$\scriptstyle{\theta_{10}}$};		
%	\node[color = black] (theta0) at (-0.6, -0.2) {$\scriptstyle{\theta_{6}}$};
%	\node[color = black] (theta1) at (-0.5, -0.8) {$\scriptstyle{\theta_{7}}$};
%	\node[color = black] (theta2) at (1.2, -0.5) {$\scriptstyle{\theta_{10}}$};		
	\node[color = black] (theta0) at (0, 0.45) {$\scriptstyle{\theta_{6}}$};
	\node[color = black] (theta1) at (0, 0.75) {$\scriptstyle{\theta_{7}}$};
	\node[color = black] (theta2) at (0, 1.05) {$\scriptstyle{\theta_{10}}$};		
%	\node (sintheta) at (1.95, 0.6) {$\scriptstyle{\sin \theta_2 }$};		
%	\node (costheta) at (0.8, -0.2) {$\scriptstyle{\cos \theta_2}$};
	\node[color = darkgray] (label1) at (2.2, 2.7) {$1$};		
	\node[color = darkgray] (labelminus1) at (-2.2, 2.7) {$-1$};		
	\node[color = darkgray] (labeli) at (2.9, 2.2) {$i$};		
	\node[color = darkgray] (labelminusi) at (2.9, -2.2) {$-i$};

%	
%	\node[inner sep = 0.4mm] (meanphase) at (-0.84072, -0.47406) {$\meanphasepos$};
	\node[inner sep = 0.4mm] (meanphase) at ({0.154508 * 2}, {-0.440838 * 2}) {$\meanphasepos$};
	\draw[->, line width = 0.3mm] (origin) -- (meanphase) {};
	
\end{tikzpicture}
\end{center}
\caption{Argand diagram of the phase positions for global state $\state_1 = \langle 0, 0, 0, 0, 0, 2, 1, 0, 0, 5 \rangle$.}

%\caption{Argand diagram of the phase positions for global state $\state = \langle 0, 1, 0, 0, 3, 0, 2, 0 \rangle$.}
\label{fig:example2}
\end{figure}

\subsection{Synchronisation Metric}
We can extend this binary notion of synchrony by introducing
a metric called \emph{phase coherence}
%with which we can 
to quantitatively measure the level of
synchrony of a global state. Our metric is derived from the
\emph{order parameter} introduced by Kuramoto \cite{kuramoto2012chemical,kuramoto1975self}
as a measure of synchrony for a population of coupled oscillators.
If we consider the phases of the oscillators as positions on the unit
circle in the complex plane then we can represent the positions
as complex numbers with magnitude $1$.
%\begin{definition}
The function $\tocomplex : \{1, \ldots, T\} \to \mathbb{C}$ maps
a phase value to its corresponding position on the unit circle in
the complex plane, and is defined as 
%$\tocomplex(\Phase) = \cos \theta_\Phase + i \sin \theta_\Phase$,
$\tocomplex(\Phase) = e^{i \theta_\Phase}$,
where  $\theta_\Phase = \frac{2 \pi}{T}(\Phase - 1)$.
%\end{definition}
A measure of synchrony $r$ can then be obtained by calculating the magnitude of
the complex number corresponding to the mean of the phase positions.
A global state has a maximal value of $r = 1$ when all oscillators
are synchronised and share the same phase $\Phase$, mapped
to the position defined by $\tocomplex(\Phase)$. It then follows that
the mean position is also $\tocomplex(\Phase)$ and 
$|\tocomplex(\Phase)| = 1$.
A global state has a minimal value of $r = 0$ when all of the positions
mapped to the phases of the oscillators are uniformly distributed
around the unit circle, or arranged  such that their positions
achieve mutual counterpoise.
%position there is another phase position
%that is the negation of the complex conjugate of  
%\begin{definition}
The \emph{phase coherence function} $\orderp : \states \to [0, 1]$
maps a global state to a real value in the interval $[0, 1]$, and is
given by
\begin{equation}
\textstyle\orderp(\langle k_1, \ldots, k_T \rangle) =
\left|
	\frac{1}{N} \sum_{\Phase = 1}^{T} k_\Phase \tocomplex(\Phase)
	\right|.
\end{equation}
%\end{definition}
Note that for any global state $\state$ where $\synch(\state)$ we
have that $\orderp(\state) = 1$, since all oscillators in $\state$
share the same phase.

%\begin{example}
Figure~\ref{fig:example2} shows a plot on the complex plane of the
positions of the phases for $N = 8$, $T = 10$, and the global state
$\state_1 = \langle 0, 0, 0, 0, 0, 2, 1, 0, 0, 5 \rangle$. The phase positions
are given by $\tocomplex(6) = e^{i \pi}$ for $2$ oscillators
with phase $6$, $\tocomplex(7) = e^{\frac{6 i \pi}{5}}$ for $1$ oscillator with phase
$7$, and $\tocomplex(10) = e^{\frac{9 i \pi}{5}}$ for $5$ oscillators with
phase $10$. We can then determine the phase coherence as %measure of synchrony for $\state$ as
$\orderp(\state) = | \frac{1}{8} (2 e^{i \pi} + e^{\frac{6i \pi}{5}} + 5 e^{\frac{9 i \pi}{5}}) | = 0.4671$.
The mean phase position is indicated on the diagram by $\meanphasepos$.
%\label{ex:orderparameter}
%\end{example}

\subsection{Correspondence with Real-Valued Oscillators}
%\pgcomment{Add an introductory sentence here explaining this}{}
Consider a clock synchronisation protocol for a cluster of $N$
fully-coupled WSN nodes, where the clocks % (values for phase)
range over real values in $[0, 2 \pi]$.
%\begin{definition}
A \emph{configuration} for $N$ oscillators
is an $N$-tuple $\conffull \in [0, 2 \pi]^N$,
where $\phasecon_j$ is the phase of oscillator $j$, for $1 \le j \le N$.
Let $\Phasecon^N$ be the set of all possible configurations for $N$ oscillators.
%\end{definition}
If the model for synchronisation can be defined as some phase response
function $\pert$, then given values for $R$, $\epsilon$, and $\mu$
we can construct a population model $\psystem$ that is a discrete
abstraction of the continuous system. When selecting a value for
$T$ there is a trade off between the size of the resulting model and
the granularity of the abstraction used to represent the oscillation
cycle.
Since the population model $\psystem$ is an abstraction of a system of
oscillators with real values for phase, and since the oscillation cycle is
represented as a sequence of $T$ discrete states, then each
discrete phase value $\Phase$ in the abstraction corresponds to
an interval of phase values $[\frac{2 \pi (\Phase{-}1)}{T}, \frac{2 \pi \Phase }{T})$
in the continuous system, having length $\frac{2 \pi}{T}$.
It then follows that if some global
state $\statefull \in \states$ is synchronised,
that is, $k_\Phase = N$ for some $1 \le \Phase \le T$,
then this corresponds to the set $\Phasecon^{N{,}T}_{\Phase}$ of possible
configurations for the oscillators in the continuous system,
where
$
\textstyle
\Phasecon^{N{,}T}_{\Phase} = \{\conffull \in \Phasecon^N  \mid \phasecon_j \in [\frac{2 \pi (\Phase{-}1)}{T}, \frac{2 \pi \Phase }{T}) \text{ for } 1 \le j \le N\}.
$
The minimum phase coherence for
all configurations in $\Phasecon^{N{,}T}_{\Phase}$ is then given by
\begin{equation}
r_{\min} = \min
	\left\{
	\textstyle
	\lvert
		\frac{1}{N}
		\sum_{j = 1}^N e^{i \theta_j}	
	\rvert
	\mid
	\conffull \in \Phasecon^{N{,}T}_{\Phase}
	\right\},
\end{equation}
where the maximum phase coherence is one, since all oscillators
may share the same phase. We therefore conclude that for some $N$, $T$,
and $\state $, if the phase coherence of the
discrete model $\orderp(\state) = 1$, then this corresponds to the phase coherence
of the continuous system being in the interval $[r_{\min}, 1]$.

\section{Model Construction}
\label{sec:model}
We use the probabilistic model checker \prism{}~\cite{kwiatkowska2011prism}
to formally verify properties of our model. Given a probabilistic model of a
system, \prism{} can be used to reason about temporal and probabilistic
properties of the input model, by checking requirements expressed in a suitable
formalism against all possible runs of the model. %This technique is known as
%\emph{model checking}.
We define our input models as
\emph{Discrete Time Markov Chains} (DTMCs).
A DTMC is a tuple $(Q, \init, \mathsf{P})$ where $Q$ is a set of states,
$\init \in Q$
is the initial state, and $\mathsf{P} : Q \times Q \rightarrow [0, 1]$ is the 
function mapping %ordered 
pairs of states $(q, q^\prime)$ to the probability with
which a transition from $q$ to $q^\prime$ occurs, where
$\sum_{q^\prime \in Q} \mathsf{P}(q, q^\prime) = 1$ for all $q \in Q$.

Given a population model $\psystemfull$ we 
construct a DTMC $D(\psystem) = (Q, \init, \mathsf{P})$. We define the set of
states $Q$ to be $\Gamma(\psystem) \cup \{\init\}$, where $\init$ is
the initial state of the DTMC. In the initial state all oscillators are %considered
%to be
\emph{unconfigured}. That is, oscillators have not yet been assigned a
value for their phase. For each
$\state = \langle k_1, \ldots, k_T \rangle \in Q \setminus \{\init\}$ we define
\begin{IEEEeqnarray}{l}
%	\mathrm{P}(\mathit{init}, q) = \frac{1}{T^N} \prod_{i = 1}^T {{N - (\sum_{j = 1}^{i  - 1} k_j )}\choose{k_i}}  \\
	\mathrm{P}(\init, q) = \frac{1}{T^N} {{N}\choose{k_1, \ldots, k_T}} %\prod_{i = 1}^T {{N - (\sum_{j = 1}^{i  - 1} k_j )}\choose{k_i}} 
\end{IEEEeqnarray}
%\begin{equation}
%	\mathsf{P}(\mathit{init}, q) = \prod_{i = 1}^T \binom{N - (\sum_{j = 1}^{i  - 1} n_j )}{n_i}
%\end{equation}
to be the probability of moving from $\init$ to a state where
$k_i$ arbitrary oscillators are configured with the phase value $i$ for
%each $i$,
$1\leq i\leq T$. The multinomial coefficient defines the number of possible
assignments of phases to distinct oscillators that result in the global state
$\state$. The fractional coefficient normalises the multinomial coefficient with
respect to the total number of possible assignments of phases to all oscillators.
In general, given an arbitrary set of initial configurations (global states)
for the oscillators, the total number of possible phase assignments can be
calculated by computing the sum of the multinomial coefficients for each
configuration (global state) in that set.  
Since $\states$
is the set of all possible global states, we have that
\begin{equation}
	\displaystyle\sum_{\langle k_1, \ldots, k_T \rangle \in \states} {{N}\choose{k_1, \ldots, k_T}} = T^N.
\end{equation}
%%%%%to be the probability of moving from $init$ to a state where the oscillators
%%%%%are \emph{configured} with the phase values defined in $q$,
%since there
%are $N$ choose $n_1$ ways to select $n_1$ oscillators to have a phase
%of $1$, then $N - n_1$ choose $n_2$ ways to select $n_2$ oscillators
%to have a phase of $2$, and so forth. 

We assign probabilities to the transitions as follows:
for every
$\state \in Q \setminus \{\init\}$ we consider each
$\state^\prime \in Q \setminus \{\init\}$ where
$\state^\prime = \suc(\state, F)$ for some
$F \in \failvecs_\state$, and set
$\mathsf{P}(\state, \state^\prime) = \pfailvec(\state, F)$. For all other
$\state\in Q \setminus \{\init\}$ and $\state^\prime \in Q$, where $\state \ne \state^\prime$
and $\state^\prime \not\in \nextstates(\state)$, we set $\mathsf{P}(\state, \state^\prime) = 0$.

To facilitate the analysis of parameterwise-different population models
we provide a Python script that allows the user to define ranges for parameters.
The script then automatically generates a model for each set of parameter values,
checks given properties in the model using \prism{},
and writes user specified output to a comma separated value file which can be
used by statistical analysis tools.\footnote{The scripts to create and analyse the data,
along with the verification results, can be found at \url{https://github.com/PaulGainer/mc-bio-synch/tree/master/energy-analysis}}

\subsection{Reward Structures}
We can annotate DTMCs with information about rewards (or costs)
by assigning values to states and transitions. By calculating the
expected value of these rewards we can reason about quantitative properties of the
models. For a network of WSN nodes we are interested in the time taken to
achieve a synchronised state and the power consumption of the network.
%A \emph{reward structure} for a DTMC is a pair $(\rstate, \rtrans)$ where
%$\rstate: Q \to \mathbb{R}^{+}$ assigns real valued rewards to states,
%and $\rtrans: Q \times Q \to \mathbb{R}^{+}$ assigns real valued
%rewards to transitions.
Given a population model $\psystemfull$, and its corresponding
DTMC $D(\psystem) = (Q, \init, \mathsf{P})$, we define the following reward
structures:

\paragraph{Synchronisation Time}
We are interested in the average and maximum time taken for a
population model to synchronise. %\uhcomment{You later define $f_s(q) =
%0$ for all $q\in Q$, not just the ones in $Q\setminus\fstates$}{To measure this we associate a reward
%with every global state $\state \in \states$ where
%$\lnot\synch(\state)$. It seems that you should delete this sentence.}
By accumulating the reward along a path until some synchronised global
state is reached we obtain a measure of the time taken to synchronise.
Recall that every global state is either a firing state or a non-firing state,
and for non-firing states we omit transitions to successor states where no
oscillators fire; instead a transition is taken to the next global state where one or
more oscillators do fire. By assigning a reward of $\frac{1}{T}$ to each transition
from each firing state, and assigning a reward of $\frac{T - \delta}{T}$ to transitions from
non-firing states to successor states, where $\delta$ is the highest phase of any oscillator in the non-firing state, and hence $T - \delta$ is the number
of omitted transitions where no oscillators fire, we obtain a measure of
synchronisation time for a population model.
%\begin{definition}
%%%Given the DTMC $D(\psystem) = (Q, init, \mathsf{P})$ for a population model $\psystem$,
%%%a \emph{synchronisation time reward structure}
%%%for $D(\psystem)$ is a tuple $\rewardsynch(\psystem) = (\rstate, \rtrans)$. We define
%%%$\rstate(q) = 0$ for all $q \in Q$. For every
%%%$(q, q^{\prime}) \in Q \times Q$, where $q \in \fstates$ and
%%%$\mathsf{P}(q, q^{\prime}) > 0$, we define
%%%$\rtrans(q, q^{\prime}) = \frac{1}{T}$. For every 
%%%$(q = \langle k_1, \ldots, k_T \rangle, q^{\prime}) \in Q \times Q$, where $q \in \nfstates$,
%%%$q^{\prime} \in \fstates$, and $\mathsf{P}(q, q^{\prime}) > 0$, we define
%%%$\rtrans(q, q^{\prime}) = \frac{T - \delta}{T}$, where
%%%$\delta = \max \{ \Phase \mid \Phase \in \{1, \ldots  T\} \land k_\Phase > 0\}$.
%%%We set $\rtrans(q, q^{\prime}) = 0$ for all other $(q, q^{\prime}) \in Q \times Q$.
%where $\mathrm{P}(q, q^{\prime}) = 0$.
%\end{definition}

\paragraph{Power Consumption}
%\begin{definition}
%A \emph{power configuration} is a tuple $(\ampdrawi, \ampdrawr, \ampdrawt, V, C, M_t)$,
Let $\ampdrawi$, $\ampdrawr$, and $\ampdrawt$  be the current draw
in amperes for the idle, receive, and transmit modes,
$V$ be the voltage, $C$ be the length of the oscillation
cycle in seconds, and $M_t$ be the time taken to transmit a
synchronisation message in seconds.
%\end{definition}
%Given a population model $\psystemfull$ %and a power configuration
%$(\ampdrawi, \ampdrawr, \ampdrawt, V, C, M_t)$, 
Let $\watthoursi = \frac{\ampdrawi V C}{3600 T}$ and
$\watthoursr = \frac{\ampdrawr V C}{3600 T}$ 
be the power consumption in Watt-hours of one node for
one discrete step within its oscillation cycle in idle and receive
mode,
and let $\watthourst  = \frac{\ampdrawt V M_t}{3600}$ be the power
consumption in Watt-hours to transmit one
synchronisation message.

The function $\power : Q \setminus \{\init\} \to \mathbb{R}$ maps a state to
the power consumption of the network in that state, given by
\begin{equation}
\power(\state) =
	\sum_{\Phase = 1}^{R} k_{\Phase} \watthoursi +
	\sum_{\Phase = R + 1}^{T} k_{\Phase} \watthoursr.
\end{equation}
The function $\powerskip : Q \cap \nfstates \to \mathbb{R}$ maps a
non-firing state to the total power consumed by the network to reach the
next firing state. Given a non firing state
$\state = \langle k_1, \ldots, k_T \rangle$
and the maximal phase
$\delta$ % = \max \{ \Phase \mid \Phase \in \{1, \ldots  T\} \land k_\Phase > 0\}$
of any oscillator in that state, we define %$\state$, we define
\begin{equation}
\powerskip(\state){=}
	\sum_{j = 0}^{(T - \delta) - 1}
	\left(
	\sum_{\Phase = 1}^{R{-}j} k_{\Phase} \watthoursi +
	\sum_{\Phase = (R{+}1){-}j}^{\delta} k_{\Phase} \watthoursr
	\right).
\label{eq:energyrewardnonfiring}
\end{equation}
From a non-firing state $\state \in Q \cap \nfstates$
the power consumed by the network to reach the next firing state
is equivalent to the accumulation of the power consumption of the
network in $\state$ and any successive non-firing states that are
omitted in the transition from $\state$ to $\sucskip(\state)$.
Furthermore, for each firing state $\state \in Q \cap \fstates$ we assign a reward of
$k_1 \watthours$ to every transition from $\state$ to a successor state
$\state^{\prime} = \langle k_1, \ldots, k_T \rangle$. This corresponds to the total
power consumption for the transmission of $k_1$ synchronisation messages.

\subsection{Restabilisation}
\label{sec:model_restab}
%We also want to reason about the synchronisation time and energy consumption for a restabilising network of oscillators. 
A network of oscillators is \emph{restabilising} if it has reached a 
synchronised state, synchrony has been lost due to the occurrence of
some external event, and the network must then again achieve synchrony.
We could, for instance, imagine the introduction of
additional nodes with arbitrary phases to an established and
synchronised network.
%\uhcomment{What follows seem to specifically address the first
%  posibility, I think it is only confusing to mention another  possibility.}{or a %hardware fault leading to the resetting of one or more nodes}.
We define the parameter $\usynch$ to be the number of oscillators with
arbitrary phase values that have been introduced into a network of $N - \usynch$
synchronised oscillators, or to be the number of oscillators in a
network of $N$ oscillators whose clocks have reset to an arbitrary value,
where
$\usynch \in \mathbb{N}$ and $1 \le \usynch < N$. 
%Note that 
Destabilising $\usynch$ oscillators in this way results in configurations
where \emph{at least} $N-\usynch$ oscillators are synchronised, since the destabilised oscillators
may coincidentally be assigned the phase of the synchronised group.
We can restrict
the set of initial configurations %for the phase values of the oscillators
by identifying the set $\ustates = \{\langle k_1, \ldots, k_T \rangle \mid \langle k_1, \ldots, k_T \rangle \in \states \text{ and } k_i \ge N - \usynch \text{ for some } 1 \le i \le T \}$, where each $\state \in \ustates$ is a configuration
for the phases such that at least $N - \usynch$ oscillators share some phase and the remaining  oscillators have arbitrary phase values.

%Since there are $N$ oscillators sharing the same phase there are then
%${\usynch + T - 2}\choose{T - 2}$ possible configurations for the remaining
%$\usynch$ oscillators. The total number of initial configurations for oscillators is then given by $T{{\usynch + %T - 2}\choose{T - 2}}$.
As we decrease the value of $\usynch$ we also decrease the number of initial configurations for the phases of the oscillators.
Since our model does not encode the loss or addition of oscillators we
can observe that all global states where there are less than $N - \usynch$ oscillators
sharing the same phase are unreachable by any run of the system beginning in some state
in $\ustates$.

\section{Evaluation}
\label{sec:eval}
%\pgcomment{I think we should probably mention the M\& S synchronisation model here, we didn't mention it at all!!!}{}
%\pgcomment{Both of us talk about 'measures of synchrony' or 'orders of %synchrony' in various places. Maybe we should be consistent with the name of %the function 'Phase Coherence Function' and refer to '(level of? measure of? %low/high?) phase coherence', or something similar. What do you think?}{}\slcomment{I think you are right. ``phase coherence'' is a better term
%than level of synchrony (and the others)}{}

In this section, we present the model checking results for instantiations
of the model given in the previous section. To that end, we instantiate the 
phase response function presented in Sect.~\ref{sec:oscillatormodel} for a 
specific synchronisation model, and vary the length of the refractory period \(R\),
coupling constant \(\epsilon\), and the probability \(\mu\) of broadcast failures.
%Note that 
All of these parameters
are global, since we assume a homogeneous network where all oscillators have identical dynamics and technical specifications.  
We use a synchronisation model 
%investigated by Lucarelli and Wang 
where
the perturbation induced by the firing of other oscillators is linear in the phase of
the perturbed oscillator and the number of firing oscillators~\cite{lucarelli2004decentralized}.
%For the synchronisation model, we choose a widely accepted definition by Mirollo and 
%Strogatz \cite{mirollo1990synchronization}, in which the oscillators are linearly perturbed according to %their current phase
%and the number of firings perceived.
That is, 
\(\pert(\Phi, \alpha, \epsilon) = \left [ \Phi {\cdot} \alpha {\cdot} \epsilon \right]\), where \([\_ ]\) denotes rounding of a value
to the nearest integer. 
The coupling constant determines the slope of the linear dependency. 

For many experiments we set \(\epsilon = 0.1\) and \(\mu = 0.2\). We could, of course,
have conducted analyses for different values for these parameters.
%We 
%For many experiments we chose \(\epsilon = 0.1\) and \(\mu = 0.2\) exemplary, however,
%a similar analysis could  of course be conducted for different parameter instantiations.
For a real system,  the probability \(\mu\) of broadcast failure occurrence is highly
dependent on the deployment environment.
For deployments in benign environments we would expect a relatively low rate of failure, for
instance a WSN within city limits under controlled conditions, whilst a comparably high
rate of failure would be expected in harsh environments such a a network of off-shore
sensors below sea level.
The coupling constant \(\epsilon\)
is a parameter of the system itself. Our results suggest that higher values
for \(\epsilon\) are always beneficial, however this is because we restrict our analysis to fully connected networks.
High values for \(\epsilon\) may be detrimental when considering different topologies, since firing nodes may perturb
synchronised subcomponents of a network. However we defer such an analysis to future work.
%several parameters: the refractory period \(R\), the coupling constant \(\epsilon\),
%and the probability \(\mu\) of broadcast failures.

%\slcomment{I didn't mean to use a different letter, but what \(T\) denotes. I read the M\& S paper by now,
%and I think ``cycle period'' is fine. }{}
%\pgcomment{'C' springs to mind, but would it get confusing (for us, not the reader) because of it being used in your concrete model for the s4 paper?}{}
As an example we analyse the power consumption of the \emph{MICAz} mote\footnote{The technical datasheet is available at \url{www.memsic.com/userfiles/files/Datasheets/WSN/micaz_datasheet-t.pdf}}. The transceiver of the MICAz mote 
possesses several modes. It can either transmit, receive, or remain idle. In transmit mode, 
it draws  \(17.4\ mA\), while in receive mode, it draws  \(19.7\ mA\). If the transceiver is idling it uses
\(20\ \mu A\)\footnote{The idle and transmit modes are composed of
several submodes. The transmit mode has three submodes 
for different transmission ranges, each of which influence the amount of current draw.  The current draw of
the idle mode depends on whether the voltage regulator is turned on or off. To account for the worst-case, we only consider submodes with
the maximal current draw.}.
%that while for example the transmit mode consists of several submodes, for different transmitting ranges, we  chose to use the maximal current
%draw for each mode.
The MICAz is powered by two AA batteries or an external power supply with a voltage of \(2.7-3.3\ V\). 
For consistency, we assume that the voltage of its power supply is \(3.0\ V\).

% \begin{table}
%   \centering
%   \caption{Current Draw of MICAz Transceiver for Different Modes }
%   \label{tab:micaz_tech}
%   \begin{tabular}{|l|l|}
% \hline
%  Idle   & \(20\ \mu A\) \\
% \hline
% Receive & \(19.7\ mA\) \\
% \hline
% Transmit & \(17.4\ mA\)\\
% \hline
%   \end{tabular}
% \end{table}

\begin{figure}
  \centering
  \includegraphics[width=\linewidth]{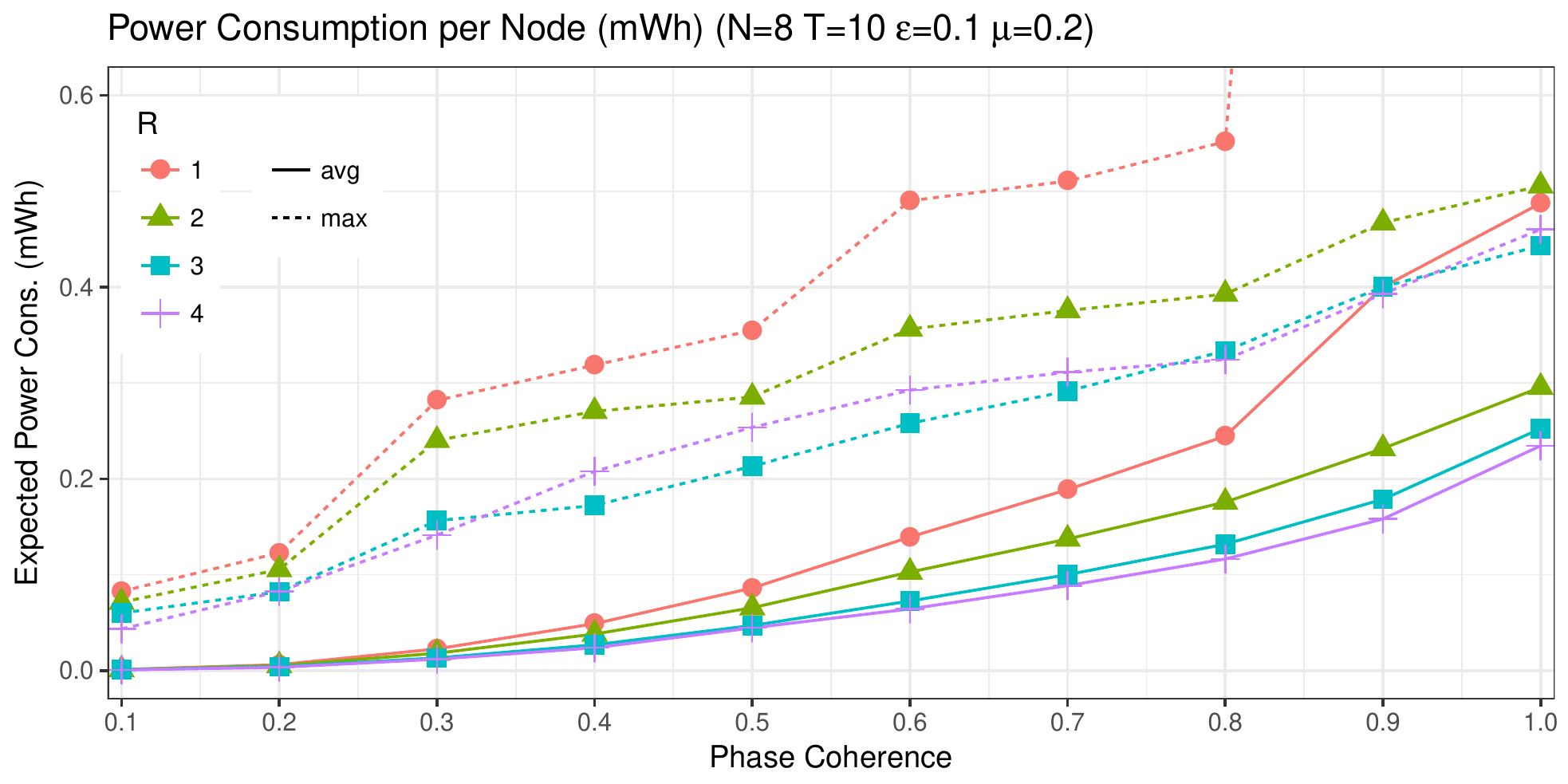}
  \caption{Power Consumption per Node to Achieve Synchronisation}
  \label{fig:pow_vs_ord}
\end{figure}

\begin{figure}
   \centering
  \includegraphics[width=\linewidth]{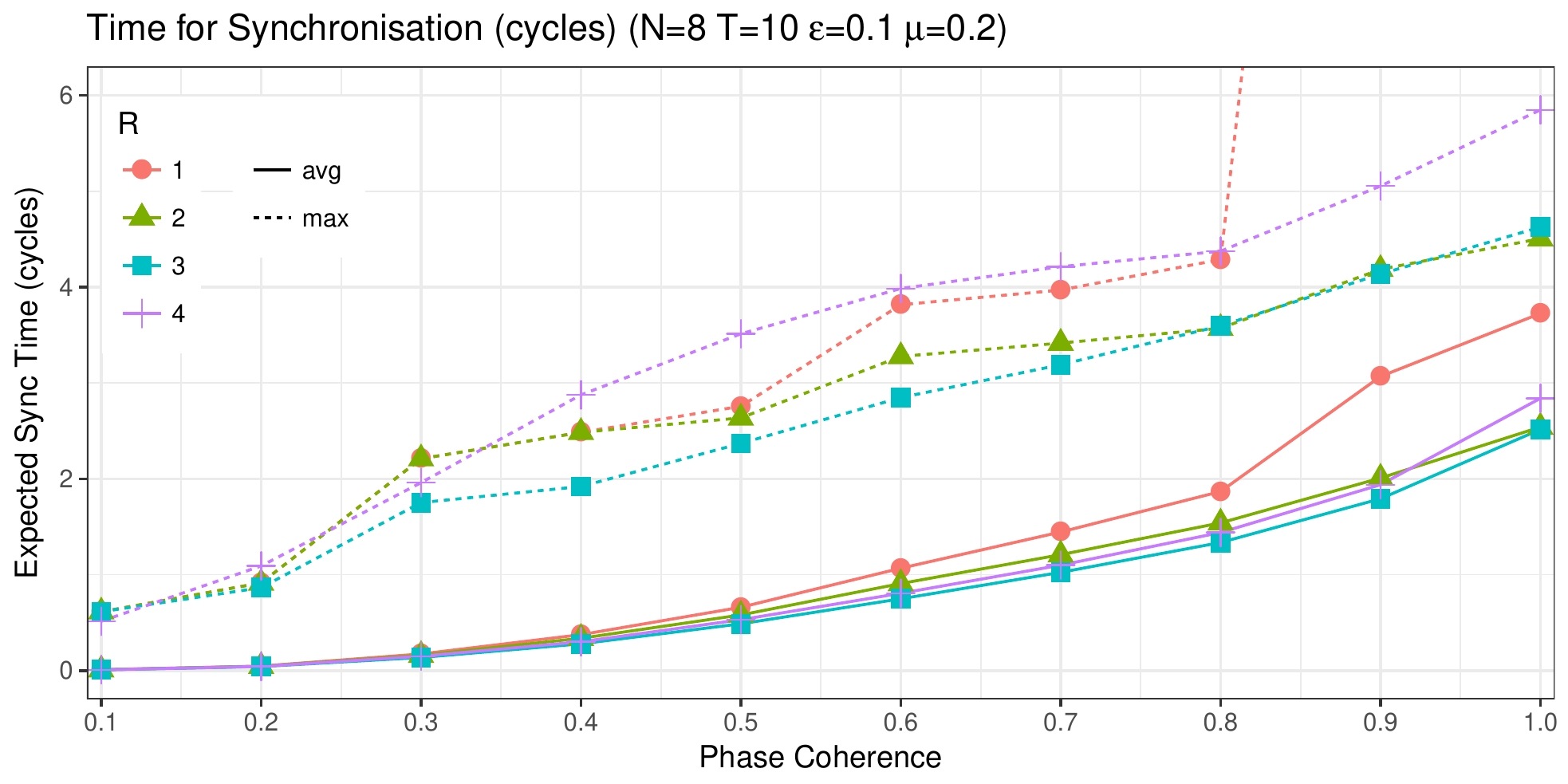}
  \caption{Time in Cycles to Achieve Synchronisation}
  \label{fig:sync_vs_ord}
\end{figure}

\subsection{Synchronisation of a whole network}
\label{sec:eval_network}
%Using these specifications
We analyse the power consumption and time to synchronise
for a network of fully connected MICAz nodes. We set the size of the network 
to be eight oscillators with a cycle period of \(T=10\). 
Increasing the granularity of the cycle period, or the size of the network, beyond these
values leads to models where it is infeasible to check properties due to time and memory
constraints\footnote{While most individual model checking runs finished within a 
minute, the cumulative model checking time over all analysed models was very large.
The results shown in Fig.~\ref{fig:pow_vs_ord} already amount to
%\(10 \cdot 4 \cdot 2 = 80\)
80 distinct runs.}.
%These values are the maximal values we were able to check, since for larger
%values the time needed for model checking became unreasonable.
However, compared to our previous work \cite{gainer2017investigating}, we were able to increase the network size.

Figures~\ref{fig:pow_vs_ord} and \ref{fig:sync_vs_ord} show both the average and maximal power consumption per node (in mWh) 
and time (in cycles) needed to synchronise,
in relation to the phase coherence of the network with respect
%\pgcomment{Do you think we need to justify why we choose these values for \(\epsilon\) and \(\mu\) here?}{}
to different lengths of the refractory period, where \(\epsilon = 0.1\) and \(\mu = 0.2\).
That is, they show
how much power is consumed (time is needed, resp.) for a system in an arbitrary state to reach a state where some degree of phase coherence has been achieved.
The much larger values obtained for \(R=1\) and phase coherence \(\geq 0.9\) are not shown here,
to avoid distortion of the figures. The energy consumption for these values is roughly \(2.4 mWh\), while the time needed is around \(19\) cycles.
Observe that we only show values for the refractory period \(R\) with \(R < \frac{T}{2}\).
For larger values of \(R\) not all runs synchronise \cite{gainer2017investigating}, resulting in an infinitely large reward being accumulated for both the maximal and average cases.
We do not provide results for the minimal power consumption (or time) as it is always zero. Since we consider all initial configurations (global states) for oscillator phases there will always be a run of the system such that the phase coherence of its initial state equals or exceeds some desired degree of phase coherence $\lambda \in [0, 1]$. This follows from the observation that for any $\lambda$  there is always an initial global state $\state$ with phase coherence $\orderp(\state) \ge \lambda$, namely any state $\state^\prime$ where all oscillators share the same phase, and hence $\orderp(\state^\prime) = 1$.
% = \langle k_1, \ldots, k_T \rangle$ where $k_i = N$ for some $1 \le  i \le T$, and $\orderp(\state) = 1 \ge \lambda$.
%As we consider \emph{all} possible initial states of the system there is always a run of the system where we begin in a state where the phase coherence of that state equals or exceeds the required degree.
%. In particular, this includes  fully synchronised initial states.  

\begin{figure}
  \centering
  \includegraphics[width=\linewidth]{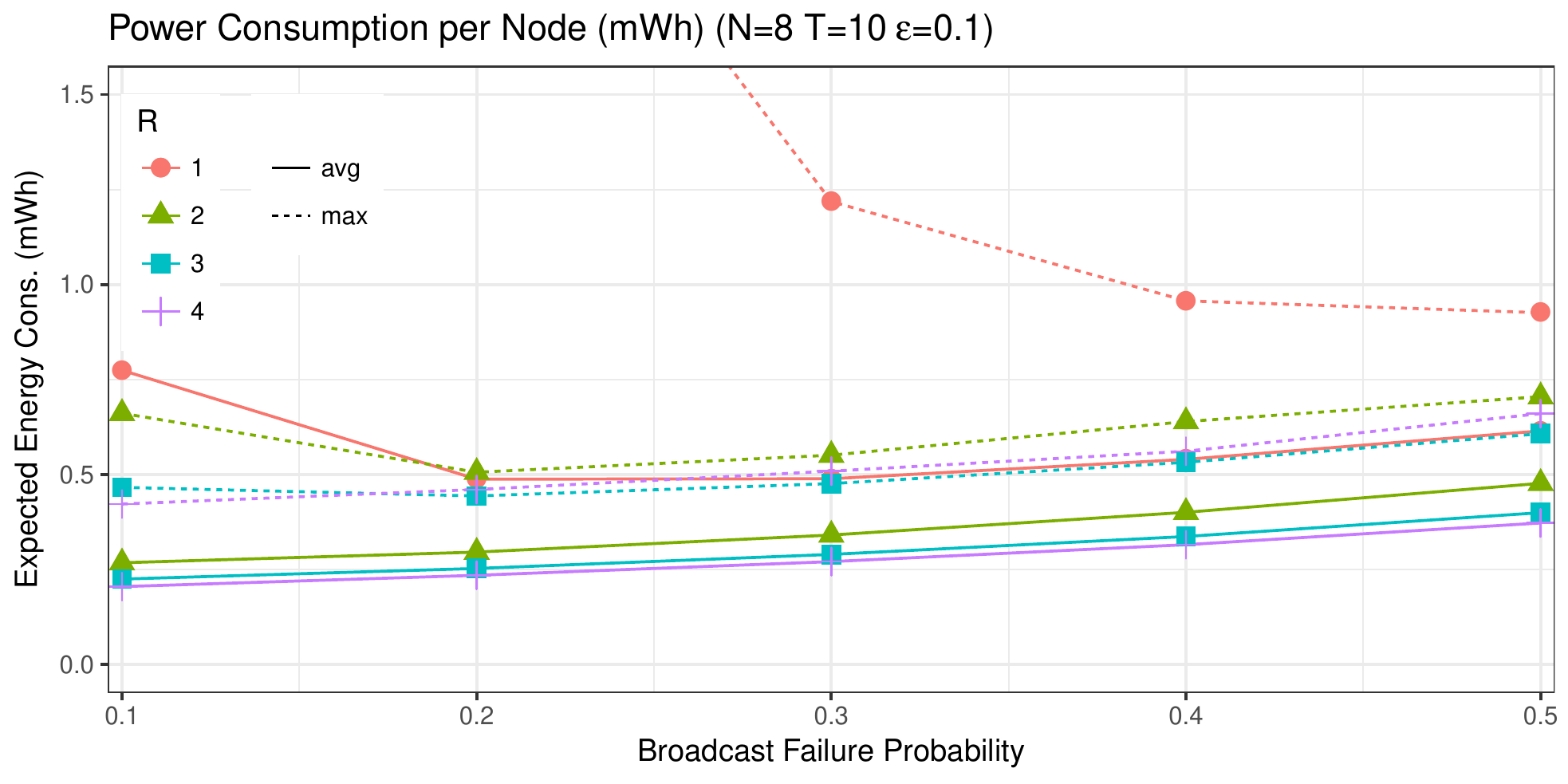}
  \caption{Power Consumption in Relation to Broadcast Failure Probability}
  \label{fig:energy_v_ml}
\end{figure}

As would be expected when starting from an arbitrary state, the expected time 
and expected power consumption increases monotonically with the order of synchrony to be achieved.
%While both variables have similar tendencies we observe a behavioural difference when considering the
% average or maximal values.
On average, networks with a higher refractory period require less 
power for synchronisation, and take less time to achieve it. The only exception
is that the average time to achieve synchrony with a refractory period of four is
higher than for two and three. However, if lower phase coherence is sufficient then
this trend is stable. In contrast to that, the maximal power consumption of networks 
with \(R=4\) is consistently higher than of networks with \(R=3\). In addition, the maximal time
needed to achieve synchrony for networks with \(R=4\) is generally higher than for 
lower refractory periods, except when the phase coherence is greater than or equal to \(0.9\). 
%For the systems with \(T=10\),
We find that networks with a refractory period of three will need the smallest amount of time to synchronise, regardless of whether we consider the maximal or
average values. Furthermore, the average power consumption for full synchronisation (phase coherence one) differs only slightly between \(R=3\) and \(R=4\) (less than 0.3 mWh). Hence, for the given example, \(R=3\) gives the best results. These relationships are stable even for different broadcast failure
probabilities  \(\mu\), while the
%\pgcomment{absolue values?}{}
concrete values increase only slightly.
This is illustrated in Fig.~\ref{fig:energy_v_ml}, which shows the average and maximal power consumption 
for different broadcast failure probabilities when \(\epsilon = 0.1\).
%In contrast, changing the coupling constant may change these relation of the behaviour with
%respect to different refractory periods. In general, however, systems with a higher refractory
%period  perform better than networks with lower refractory periods.  

\begin{figure}
  \centering
  \includegraphics[width=\linewidth]{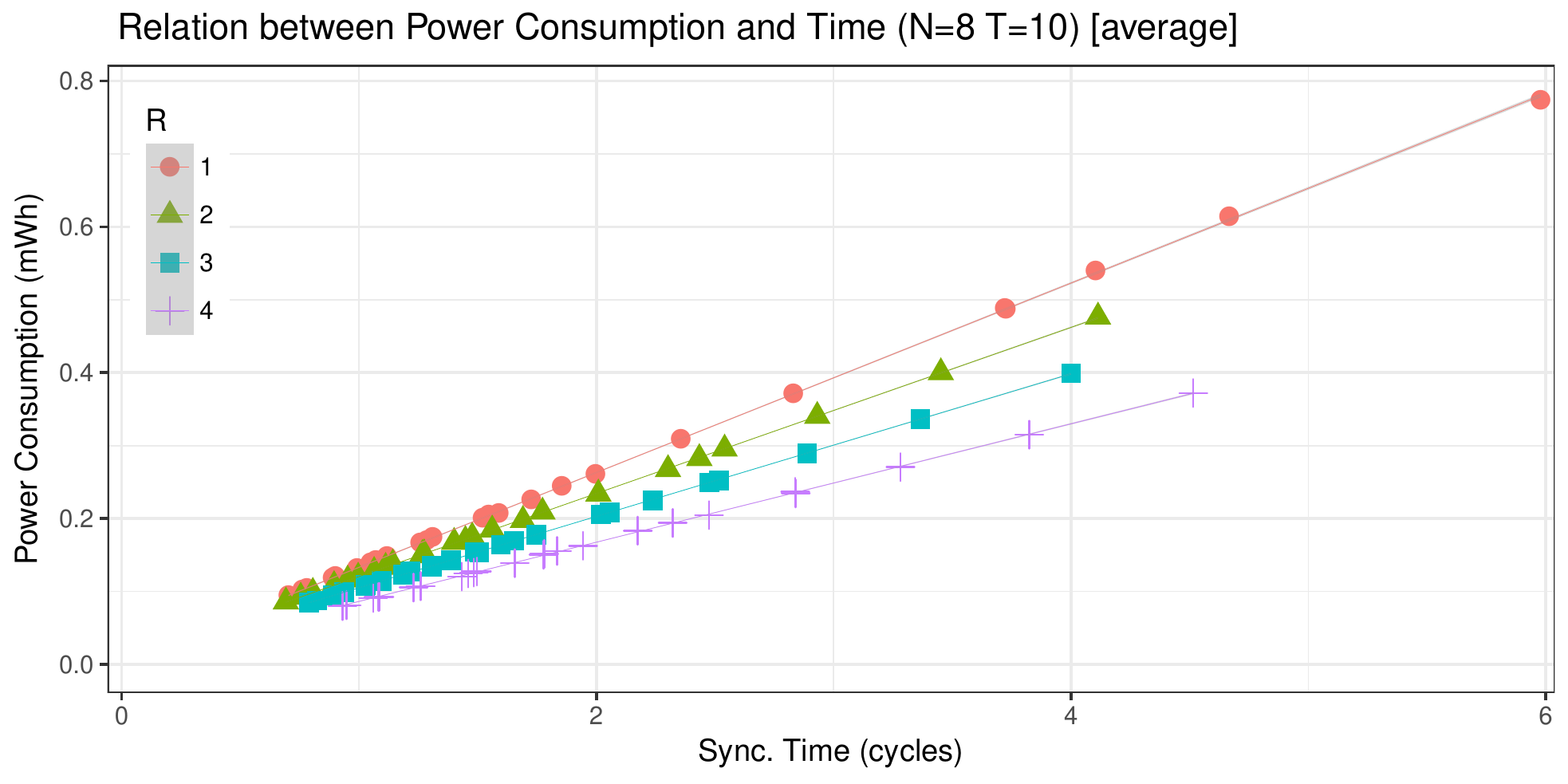}
  \caption{Average Power Consumption to Average Time for Synchrony}
  \label{fig:energy_v_time_avg}
\end{figure}

\begin{figure}[t]
  \centering
  \includegraphics[width=\linewidth]{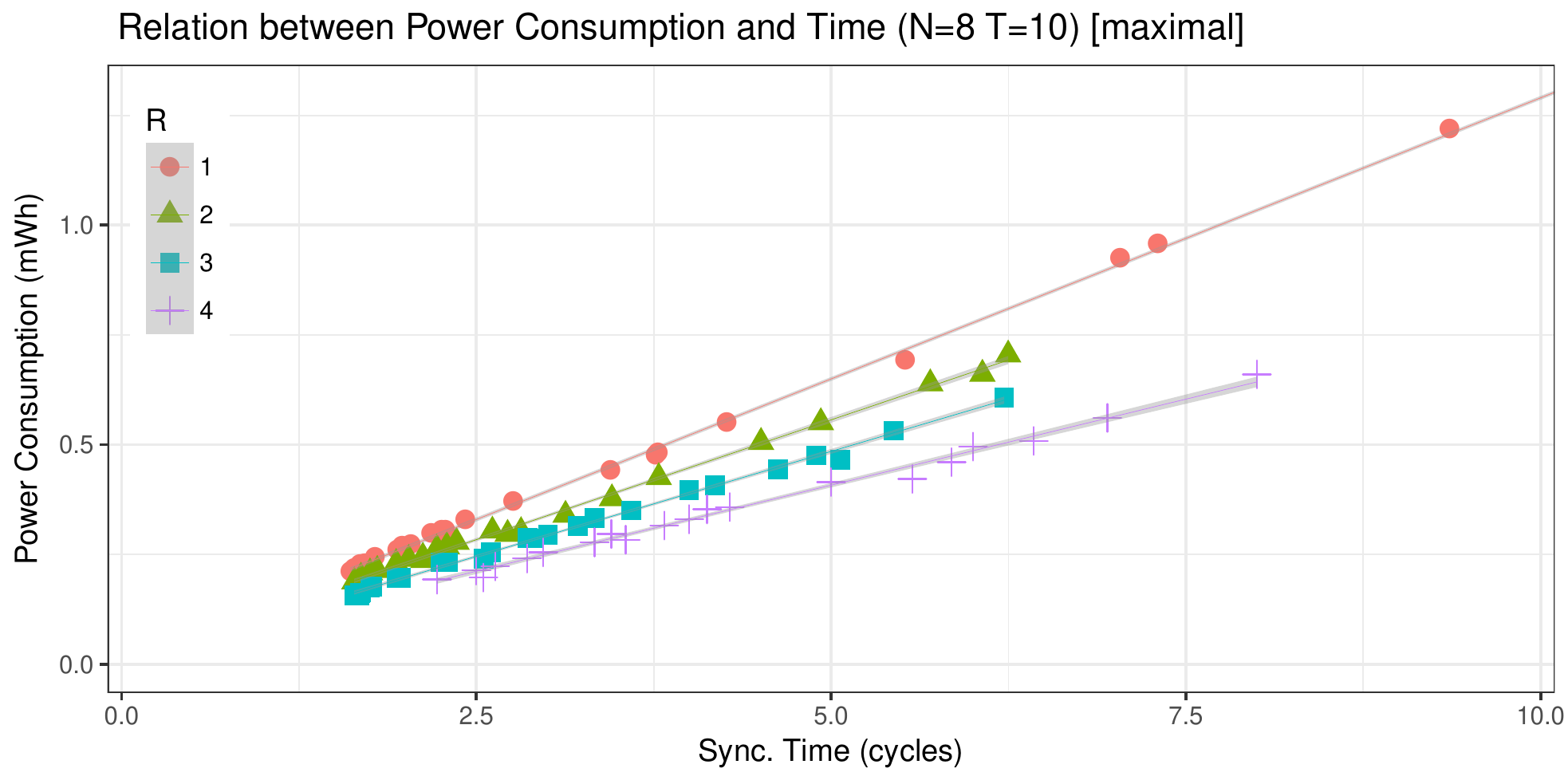}
  \caption{Maximal Power Consumption  to Maximal Time for Synchrony}
  \label{fig:energy_v_time_max}
\end{figure}

The general relationship between power consumption and time needed to synchronise is 
shown in Figs.~\ref{fig:energy_v_time_avg} and \ref{fig:energy_v_time_max}. Within these figures,
we do not distinguish between different coupling constants and broadcast failure
probabilities.
We omit the two values for \(R=1\), \(\epsilon = 0.1\) and \(\mu \in \{0.1, 0.2\}\) in Fig.~\ref{fig:energy_v_time_max} to avoid distortion of the graph,
since the low coupling strength and low probability of broadcast failure leads to longer
synchronisation times and hence higher power consumption. While this might seem surprising
it has been shown that uncertainty in discrete systems often aids convergence~\cite{gainer2017investigating,fates2015remarks}.

%Also note that we do not show two values for \(R=1\) in Fig.~\ref{fig:energy_v_time_max}, since
%they are far larger than the other results. These values are for \(\epsilon = 0.1\) and \(\mu \in \{0.1, 0.2\}\), i.e.,
%very light coupling and a low probability of broadcast failures.

It is easy to see that the relationship between power consumption and time to synchronise is linear, and that the slope of the relation decreases
for higher refractory periods. While the linearity of the relation
is almost perfect for the average values, the maximal values have greater variation.
These relationships again suggest that \(R=3\) is a sensible and reliable choice for the
length of refractory period, since it provides the greatest stability
%\pgcomment{Stability seems weird, need a better description here.}{}
of power consumption and time
to synchronise. In particular, if the broadcast failure probability changes, the variations
in power consumption and synchronisation time are less severe for \(R=3\) than for
the other refractory period lengths.

\subsection{Resynchronisation of a small number of nodes}
\label{sec:eval_resync}

%\uhcomment{The first two sentences seem to repeat what was said in
%  Sect.~\ref{sec:model_restab}. I suggest to delete these two
%  sentences.}{After the deployment of a network of synchronising nodes, oscillators may be replaced or may temporarily lose
%their connection to the rest of the network. This may result in a situation where these oscillators
%are no longer in synchrony with the other nodes.}
In this section, we present an analysis of 
the power consumption if the number of redeployed nodes is small compared to the  
size of the network. The approach presented in Sect.~\ref{sec:model_restab} allows us to 
significantly increase the network size. % compared with the previous section.
In particular,
the smallest network we analyse is already larger than that in Sect.~\ref{sec:eval_network}, while the largest  is almost five times as large. This is possible because the model checker only needs to construct a much smaller number of initial states. % than in the general case.

\begin{figure}[t]
  \centering
  \includegraphics[width=\linewidth]{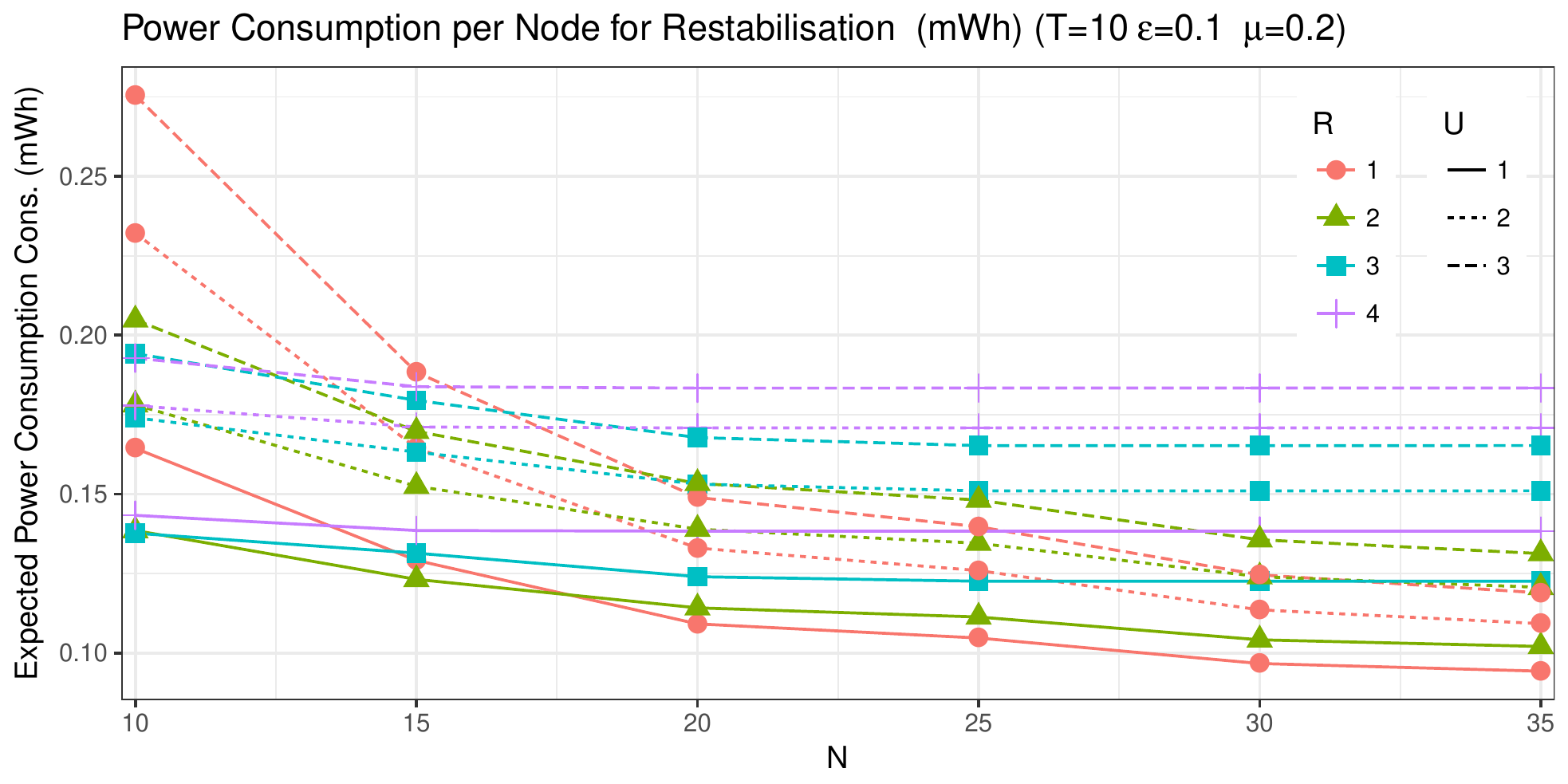}
  \caption{Average Power Consumption for Resynchronisation to Network Size}
  \label{fig:energy_v_N}
\end{figure}

%The model checking results 
The average power consumption per node
for networks of size \(10, 15, \ldots, 35\), where
the oscillators are coupled with strength \(\epsilon = 0.1\) and the probability of
broadcast failures is \(\mu = 0.2\) is shown in Fig.~\ref{fig:energy_v_N}. The solid lines
denote the results for a single redeployed node, while the dashed lines
represent the results for the redeployment of two and three nodes, respectively.
%Again, the results show the power consumption on average \emph{per node}.
As expected,  the more nodes need to resynchronise, the more energy is consumed. 
However, we can also extract that for higher refractory periods, the amount of energy needed
is more or less stable, in particular, in case $R=4$,
which is already invariant for more than ten nodes. 
For smaller refractory periods, % we can see that by 
increasing the network size,
decreases the average energy consumption.
% generally decreases. 
 This behaviour can be explained as follows. 
%For one, the relative size of the set of resynchronising oscillators decreases, compared
%to the network size. Furthermore, 
The linear synchronisation model implies that
oscillators with a higher phase value will be activated more and thus are more likely to
fire. For example, consider a network of \(N=15\) oscillators with a cycle length of \(T=10\)
and a coupling constant of \(\epsilon =0.1\), where one oscillator has to
resynchronise. Furthermore, assume 
that the \(14\) synchronised oscillators have phase ten and that the single oscillator has just
left its refractory period. If \(R=1\), then this means the single oscillator has a phase of two, 
and the perturbation is \([2 \cdot 14 \cdot 0.1] = [2.8] = 3\). Hence, it will not synchronise
with the other oscillators in this cycle. However, if \(R=4\), then the perturbation is \([5 \cdot 14 \cdot 0.1] =  7 \),
which is large enough to let the oscillator fire as well, i.e., it is absorbed by rest of the
oscillators and hence synchronised.
%In particular, this means, of course, that in general,  a larger network will force the node to 
This means that in general a larger network will force the node to 
resynchronise faster. The refractory period determines how large the network has to be for this effect
to stabilise.

%if the refractory period is smaller, the probability that
%the de-synchronised oscillators fire and perturb the group of synchronised nodes is higher, which
%may postpone the synchronisation of the whole network.\slcomment{Does this make sense? Maybe an example?}{}
%However, for increasing the sizes of the network, a higher
%refractory period implies a higher amount of energy consumption per node.  

%%% Local Variables: 
%%% mode: latex
%%% TeX-master: "main"
%%% End: 
\section{Conclusion}
\label{sec:conc}

We presented a formal model to analyse power consumption
in fully connected networks of PCOs. %pulse-coupled oscillators.
To that end,
we extended an existing model for synchrony convergence with a reward structure to reflect
the energy consumption of wireless sensor nodes.
Furthermore, we showed how to mitigate the state-space explosion typically encountered when model-checking. In particular, 
the state space can be reduced by ignoring states where there are no interactions
between oscillators.
When investigating the restabilisation of a small number of oscillators in an already
synchronised network we can reduce the state space significantly, since only a small subset
of the initial states needs to be considered. We used these techniques to analyse
the power consumption for synchronisation and restabilisation of a network of MICAz motes,
using the pulse-coupled oscillator model developed by Mirollo and Strogatz \cite{mirollo1990synchronization} with a 
linear phase response function. By using our model we were able to extend the
size of the network compared with previous work \cite{gainer2017investigating} and 
discuss trade-offs between the time and power needed to synchronise for different 
lengths of the refractory period (or duty cycle).

Results obtained using these techniques can be used by designers of WSNs
to estimate the overall energy efficiency of a network during its design phase.
That is, unnecessary energy consumption can be identified and rectified before
deployment of the network. Additionally, our results provide guidance for
estimating the battery life expectancy of a network depending on the anticipated 
frequency of restabilisations. Of course, these considerations 
only hold for the maintenance task of synchronising the network. The energy consumption
of the functional behaviour has to be examined separately.

Our current approach is restricted to fully connected networks of oscillators. While
this is sufficient to analyse the behaviour of strongly connected components within a 
network, further investigation is needed to assess the effect of different topological
properties on the network.
To that end, 
we could use several interconnected population models
thus modelling the interactions of the networks subcomponents.
Furthermore, topologies that change over time are of particular interest.
However, it is not obvious how we could extend our approach to consider such dynamic
networks. The work of Lucarelli and Wang may serve as a starting point
for further investigations \cite{lucarelli2004decentralized}. 
Stochastic node failure, as well as more subtle models of energy
consumption, present significant opportunities for future
extensions. For example, in some cases, repeatedly powering nodes
on and off over short periods of time might use considerably more power than
leaving them on throughout.
%%% Local Variables: 
%%% mode: latex
%%% TeX-master: "main"
%%% End: 

\section*{Acknowledgment}
This work was supported by both the Sir Joseph Rotblat Alumni Scholarship at Liverpool
and the EPSRC Research Programme EP/N007565/1 \emph{Science of Sensor Systems Software}.
The authors would like to thank the Networks Sciences and Technology Initiative
(NeST) of the University of Liverpool for the use of their computing facilities and
David Shield for the corresponding technical support.

\bibliographystyle{IEEEtran}
\bibliography{litreview}

\begin{thebibliography}{10}
\providecommand{\url}[1]{#1}
\csname url@samestyle\endcsname
\providecommand{\newblock}{\relax}
\providecommand{\bibinfo}[2]{#2}
\providecommand{\BIBentrySTDinterwordspacing}{\spaceskip=0pt\relax}
\providecommand{\BIBentryALTinterwordstretchfactor}{4}
\providecommand{\BIBentryALTinterwordspacing}{\spaceskip=\fontdimen2\font plus
\BIBentryALTinterwordstretchfactor\fontdimen3\font minus
  \fontdimen4\font\relax}
\providecommand{\BIBforeignlanguage}[2]{{%
\expandafter\ifx\csname l@#1\endcsname\relax
\typeout{** WARNING: IEEEtran.bst: No hyphenation pattern has been}%
\typeout{** loaded for the language `#1'. Using the pattern for}%
\typeout{** the default language instead.}%
\else
\language=\csname l@#1\endcsname
\fi
#2}}
\providecommand{\BIBdecl}{\relax}
\BIBdecl

\bibitem{rhee2004techniques}
S.~Rhee, D.~Seetharam, and S.~Liu, ``Techniques for minimizing power
  consumption in low data-rate wireless sensor networks,'' in \emph{Wireless
  Communications and Networking Conference, 2004. WCNC. 2004 IEEE},
  vol.~3.\hskip 1em plus 0.5em minus 0.4em\relax IEEE, 2004, pp. 1727--1731.

\bibitem{albers2010energy}
S.~Albers, ``Energy-efficient algorithms,'' \emph{Communications of the ACM},
  vol.~53, no.~5, pp. 86--96, 2010.

\bibitem{irani2005algorithmic}
S.~Irani and K.~R. Pruhs, ``Algorithmic problems in power management,''
  \emph{ACM Sigact News}, vol.~36, no.~2, pp. 63--76, 2005.

\bibitem{soua2011survey}
R.~Soua and P.~Minet, ``A survey on energy efficient techniques in wireless
  sensor networks,'' in \emph{Wireless and Mobile Networking Conference (WMNC),
  2011 4th Joint IFIP}.\hskip 1em plus 0.5em minus 0.4em\relax IEEE, 2011, pp.
  1--9.

\bibitem{oller2016has}
J.~Oller, I.~Demirkol, J.~Casademont, J.~Paradells, G.~U. Gamm, and L.~Reindl,
  ``Has time come to switch from duty-cycled mac protocols to wake-up radio for
  wireless sensor networks?'' \emph{IEEE/ACM Transactions on Networking},
  vol.~24, no.~2, pp. 674--687, 2016.

\bibitem{yick2008survey}
\BIBentryALTinterwordspacing
J.~Yick, B.~Mukherjee, and D.~Ghosal, ``Wireless sensor network survey,''
  \emph{Computer Networks}, vol.~52, no.~12, pp. 2292 -- 2330, 2008. [Online].
  Available:
  \url{http://www.sciencedirect.com/science/article/pii/S1389128608001254}
\BIBentrySTDinterwordspacing

\bibitem{rhee2009clock}
I.-K. Rhee, J.~Lee, J.~Kim, E.~Serpedin, and Y.-C. Wu, ``Clock synchronization
  in wireless sensor networks: An overview,'' \emph{Sensors}, vol.~9, no.~1,
  pp. 56--85, 2009.

\bibitem{taniguchi2005distributed}
Y.~Taniguchi, N.~Wakamiya, and M.~Murata, ``A distributed and self-organizing
  data gathering scheme in wireless sensor networks,'' in \emph{6th
  Asia-Pacific Symposium on Information and Telecommunication
  Technologies}.\hskip 1em plus 0.5em minus 0.4em\relax IEEE, 2005, pp.
  299--304.

\bibitem{tyrrell2006fireflies}
A.~Tyrrell, G.~Auer, and C.~Bettstetter, ``Fireflies as role models for
  synchronization in ad hoc networks,'' in \emph{Proceedings of the 1st
  international conference on Bio inspired models of network, information and
  computing systems}.\hskip 1em plus 0.5em minus 0.4em\relax ACM, 2006, p.~4.

\bibitem{bojic2014scalability}
I.~Bojic, T.~Lipic, and M.~Kusek, ``Scalability issues of firefly-based
  self-synchronization in collective adaptive systems,'' in \emph{Proc. SASOW
  2014}.\hskip 1em plus 0.5em minus 0.4em\relax IEEE, 2014, pp. 68--73.

\bibitem{lipa2015firefly}
N.~Lipa, E.~Mannes, A.~Santos, and M.~Nogueira, ``Firefly-inspired and robust
  time synchronization for cognitive radio ad hoc networks,'' \emph{Computer
  Communications}, vol.~66, pp. 36--44, 2015.

\bibitem{werner2005firefly}
G.~Werner-Allen, G.~Tewari, A.~Patel, M.~Welsh, and R.~Nagpal,
  ``Firefly-inspired sensor network synchronicity with realistic radio
  effects,'' in \emph{Proc. SenSys 2005}.\hskip 1em plus 0.5em minus
  0.4em\relax ACM, 2005, pp. 142--153.

\bibitem{peskin1975mathematical}
C.~Peskin, \emph{Mathematical aspects of heart physiology}, ser. Courant
  Lecture Notes.\hskip 1em plus 0.5em minus 0.4em\relax Courant Institute of
  Mathematical Sciences, New York University, 1975.

\bibitem{mirollo1990synchronization}
R.~E. Mirollo and S.~H. Strogatz, ``Synchronization of pulse-coupled biological
  oscillators,'' \emph{SIAM J. App. Math.}, vol.~50, no.~6, pp. 1645--1662,
  1990.

\bibitem{lucarelli2004decentralized}
D.~Lucarelli, I.-J. Wang \emph{et~al.}, ``Decentralized synchronization
  protocols with nearest neighbor communication,'' in \emph{Proc. SenSys
  2004}.\hskip 1em plus 0.5em minus 0.4em\relax ACM, 2004, pp. 62--68.

\bibitem{kwiatkowska2007stochastic}
M.~Kwiatkowska, G.~Norman, and D.~Parker, ``Stochastic model checking,'' in
  \emph{SFM}, vol.~7.\hskip 1em plus 0.5em minus 0.4em\relax Springer, 2007,
  pp. 220--270.

\bibitem{baier2010performability}
C.~Baier, L.~Cloth, B.~R. Haverkort, H.~Hermanns, and J.-P. Katoen,
  ``Performability assessment by model checking of markov reward models,''
  \emph{Formal Methods in System Design}, vol.~36, no.~1, pp. 1--36, 2010.

\bibitem{baier2014trade}
C.~Baier, C.~Dubslaff, and S.~Kl{\"u}ppelholz, ``Trade-off analysis meets
  probabilistic model checking,'' in \emph{Proceedings of the Joint Meeting of
  the Twenty-Third EACSL Annual Conference on Computer Science Logic (CSL) and
  the Twenty-Ninth Annual ACM/IEEE Symposium on Logic in Computer Science
  (LICS)}.\hskip 1em plus 0.5em minus 0.4em\relax ACM, 2014, p.~1.

\bibitem{chen2013review}
Z.~Chen, D.~Zhang, R.~Zhu, Y.~Ma, P.~Yin, and F.~Xie, ``A review of automated
  formal verification of ad hoc routing protocols for wireless sensor
  networks,'' \emph{Sensor Letters}, vol.~11, no.~5, pp. 752--764, 2013.

\bibitem{fehnker2006formal}
A.~Fehnker and P.~Gao, ``Formal verification and simulation for performance
  analysis for probabilistic broadcast protocols,'' in \emph{Proc. ADHOC-NOW
  2006}, ser. LNCS, vol. 4104.\hskip 1em plus 0.5em minus 0.4em\relax Springer,
  2006, pp. 128--141.

\bibitem{yue2010leader}
H.~Yue and J.-P. Katoen, ``Leader election in anonymous radio networks: Model
  checking energy consumption.'' in \emph{ASMTA}.\hskip 1em plus 0.5em minus
  0.4em\relax Springer, 2010, pp. 247--261.

\bibitem{heidarian2012analysis}
F.~Heidarian, J.~Schmaltz, and F.~Vaandrager, ``Analysis of a clock
  synchronization protocol for wireless sensor networks,'' \emph{Theor. Comput.
  Sci.}, vol. 413, no.~1, pp. 87--105, 2012.

\bibitem{bartocci2010detecting}
E.~Bartocci, F.~Corradini, E.~Merelli, and L.~Tesei, ``Detecting
  synchronisation of biological oscillators by model checking,'' \emph{Theor.
  Comput. Sci.}, vol. 411, no.~20, pp. 1999--2018, 2010.

\bibitem{alur1994theory}
R.~Alur and D.~L. Dill, ``A theory of timed automata,'' \emph{Theor. Comput.
  Sci.}, vol. 126, no.~2, pp. 183--235, 1994.

\bibitem{gainer2017investigating}
P.~Gainer, S.~Linker, C.~Dixon, U.~Hustadt, and M.~Fisher, ``Investigating
  parametric influence on discrete synchronisation protocols using quantitative
  model checking,'' in \emph{Proc. QEST 2017}, ser. LNCS.\hskip 1em plus 0.5em
  minus 0.4em\relax Springer, 2017.

\bibitem{wang2012energy}
Y.~Wang, F.~Nu\~{n}ez, and F.~J. Doyle, ``Energy-efficient pulse-coupled
  synchronization strategy design for wireless sensor networks through reduced
  idle listening,'' \emph{IEEE Transactions on Signal Processing}, vol.~60,
  no.~10, pp. 5293--5306, 2012.

\bibitem{konishi2008synchronization}
\BIBentryALTinterwordspacing
K.~Konishi and H.~Kokame, ``Synchronization of pulse-coupled oscillators with a
  refractory period and frequency distribution for a wireless sensor network,''
  \emph{Chaos: An Interdisciplinary Journal of Nonlinear Science}, vol.~18,
  no.~3, 2008. [Online]. Available:
  \url{http://aip.scitation.org/doi/abs/10.1063/1.2970103}
\BIBentrySTDinterwordspacing

\bibitem{emerson1999asymmetry}
E.~A. Emerson and R.~J. Trefler, ``From asymmetry to full symmetry: New
  techniques for symmetry reduction in model checking.'' in \emph{CHARME},
  vol.~99.\hskip 1em plus 0.5em minus 0.4em\relax Springer, 1999, pp. 142--156.

\bibitem{donaldson2006symmetry}
A.~F. Donaldson and A.~Miller, ``Symmetry reduction for probabilistic model
  checking using generic representatives,'' in \emph{ATVA}, vol.~6.\hskip 1em
  plus 0.5em minus 0.4em\relax Springer, 2006, pp. 9--23.

\bibitem{gainer2016probabilistic}
P.~Gainer, C.~Dixon, and U.~Hustadt, ``Probabilistic model checking of
  ant-based positionless swarming,'' in \emph{Proc. TAROS 2016}, ser. LNCS,
  vol. 9716.\hskip 1em plus 0.5em minus 0.4em\relax Springer, 2016, pp.
  127--138.

\bibitem{kwiatkowska2011prism}
M.~Kwiatkowska, G.~Norman, and D.~Parker, ``Prism 4.0: Verification of
  probabilistic real-time systems,'' in \emph{Proc. CAV 2011}, ser. LNCS, vol.
  6806.\hskip 1em plus 0.5em minus 0.4em\relax Springer, 2011, pp. 585--591.

\bibitem{hull2004mitigating}
B.~Hull, K.~Jamieson, and H.~Balakrishnan, ``Mitigating congestion in wireless
  sensor networks,'' in \emph{Proc. SenSys 2004}.\hskip 1em plus 0.5em minus
  0.4em\relax ACM, 2004, pp. 134--147.

\bibitem{kuramoto2012chemical}
Y.~Kuramoto, \emph{Chemical oscillations, waves, and turbulence}.\hskip 1em
  plus 0.5em minus 0.4em\relax Springer Science \& Business Media, 2012,
  vol.~19.

\bibitem{kuramoto1975self}
------, ``Self-entrainment of a population of coupled non-linear oscillators,''
  in \emph{International Symposium on Mathematical Problems in Theoretical
  Physics}, ser. LNP, vol.~39.\hskip 1em plus 0.5em minus 0.4em\relax Springer,
  1975, pp. 420--422.

\bibitem{kramer2006energy}
M.~Kramer and A.~Geraldy, ``Energy measurements for {MICA}z node,''
  \emph{University of Kaiserslautern, Kaiserslautern, Germany, Technical Report
  KrGe06}, 2006.

\bibitem{webster2017performance}
M.~Webster, M.~Breza, C.~Dixon, M.~Fisher, and J.~McCann, ``Performance
  evaluation of gossip{-}synchronization algorithms for wireless sensor
  networks using formal verification,'' {S}ubmitted for publication.

\bibitem{fates2015remarks}
N.~Fat{\`e}s, ``Remarks on the cellular automaton global synchronisation
  problem,'' in \emph{Proc. AUTOMATA 2015}, ser. LNCS, vol. 9099.\hskip 1em
  plus 0.5em minus 0.4em\relax Springer, 2015, pp. 113--126.

\end{thebibliography}

\end{document}